\documentclass[prl,aps,preprintnumbers,twocolumn, acknowledgement]{revtex4}

\usepackage{amsmath,amssymb,bm}
\usepackage{graphicx} 
\usepackage{color}

\def\eqa{\begin{eqnarray}}
\def\eea{\end{eqnarray}}
\newcommand{\eq}{\begin{equation}}
\newcommand{\ee}{\end{equation}}

\newcommand{\Tr}{{\rm Tr}}

\newcommand{\ket}[1]{\ensuremath{\left|#1\rangle \right.}}
\newcommand{\bra}[1]{\ensuremath{\left. \langle #1 \right|}}
\newcommand\inner[2]{\ensuremath{\langle #1  | #2  \rangle }}
\newcommand\dirac[3]{\ensuremath{\langle #1 | #2 | #3 \rangle }}

\begin{document}
\title{Quantum Theory of Phonon Induced Anomalous Hall Effect in 2D Massive Dirac metals}
\author{Jia-Xing Zhang$^{1}$}
\author{Wei Chen$^{1,2}$} \email{chenweiphy@nju.edu.cn}
\affiliation{$^{1}$National Laboratory of Solid State Microstructures and School of Physics, Nanjing University, Nanjing, China}
\affiliation{$^2$Collaborative Innovation Center of Advanced Microstructures, Nanjing University, Nanjing, China}

\begin{abstract}
The phonon induced anomalous Hall or thermal Hall effects have been observed in various systems in recent experiments. However, the theoretical studies on this subject are still incomplete and not unified, and the current works mainly focus on the semi-classical Boltzmann equation approach. In this work, we present a  systematic and fundamental quantum field theory study on the phonon induced anomalous Hall effect, including both the side jump and skew scattering contributions, in a 2D massive Dirac metal, which is considered as the minimum anomalous Hall system. We reveal significant difference from the anomalous Hall effect induced by the widely studied Gaussian disorder which is known to be insensitive to temperature. While the anomalous Hall effect induced by phonon (deformation potential) approaches that by Gaussian disorder at high temperature, it behaves very differently at low temperature. Our work provides a microscopic and quantitative description of the crossover from the low to high temperature regime of the phonon induced anomalous Hall conductivity, which may be observed in 2D Dirac metals with breaking time reversal symmetry.
\end{abstract}

\maketitle  

The anomalous Hall effect (AHE), a transverse voltage arising in a metal or semiconductor in response to an applied current without magnetic field, was experimentally discovered as early as 1881~\cite{Hall1881}. Ever since then, the search for the microscopic origin of the AHE has been one of the main issues of condensed matter physics~\cite{Smit1955, Luttinger1955, Luttinger1958, Luttinger1964, Berger1970, Streda1982, Sinitsyn2006, Sinitsyn2007, Sinitsyn2008, Shindou2006, Nagaosa2010}. The  subsequent discoveries  have brought the important aspect of topology to modern condensed matter physics~\cite{Haldane2004, Niu2010} and led  to many important applications. 
After a long-lasting debate, it is established that there are two types of mechanisms which may result in AHE in materials with broken time reversal symmetry (TRS)~\cite{Sinitsyn2007, Sinitsyn2008, Nagaosa2010}: One is the intrinsic contribution which comes from the nontrivial Berry curvature of the band structure; 
the other is the extrinsic contribution which originates from electron scatterings by impurities in materials with (pseudo-)spin-orbit interaction. The latter can  be further divided to the side jump contribution, which is due to transverse coordinate shift by scatterings, and the skew scattering contributions, which is  due to asymmetric scatterings~\cite{Sinitsyn2007}.

Most previous studies on the extrinsic contribution have focused on electron scatterings off static disorder~\cite{Sinitsyn2006, Sinitsyn2007, Sinitsyn2008, Nagaosa2010, MacDonald2006, Yang2011}. However, recent experiments have detected various anomalous Hall or thermal Hall effects dominated by electron scatterings off phonons~\cite{Yao2022, Sharma2024}. Yet the theoretical studies on the phonon-induced AHE are still incomplete and not unified. Recent theoretical works on this subject mainly focus on the semi-classical Boltzmann equation (SBE) approach.
In ~\cite{Niu2019}, the authors present a SBE approach  for the phonon induced side jump conductivity in the 2D massive Dirac model with a justification by the Argyres-Kohn-Luttinger and Lyo-Holstein quantum transport theory. In a following work~\cite{Xiao2019-1}, the authors further applied the SBE approach to the phonon-induced intrinsic skew scattering contribution (which comes from non-crossing diagrams)  and studied the scaling parameters between the anomalous Hall and longitudinal resistivity based on this approach.
In a later work by another group~\cite{Glazov2020}, the authors studied a related effect, the valley Hall effect in a 2D massive Dirac model in the high temperature regime due to phonon drag and phonon  scatterings  with the same semi-classical approach verified by the Keldysh technique. Due to the different  formulation and approximations taken in the semi-classical approach, the semi-classical results from different groups often have different forms.
Besides, the skew scattering contribution from the crossed diagrams due to phonon scatterings has not been studied quantitatively in previous work.
For these reasons, a fundamental and systematic quantum field theory (QFT) study of the phonon induced AHE  is valuable and a good test of the different semi-classical works, and this is what we do in this work.

In this paper, we present a systematic QFT study of the AHE induced  by phonon scatterings in a 2D massive Dirac metal~\cite{Sinitsyn2007, MacDonald2006}, including both the contribution from the non-crossing and crossed Feynman diagrams. 
For simplicity, we focus on the scalar phonon mode, or the deformation potential (DP) induced AHE in this system. We obtained the AH conductivity, including both the side jump and skew scattering contributions due to phonon scatterings in the temperature range $T\ll \epsilon_F$, as plotted in Fig.\ref{fig:AH_conductivity}. The analytic results of the AH conductivities in the limit $T\ll T_{BG}$ and $T\gg T_{BG}$, where $T_{BG}\equiv 2s k_F$ is the Bloch-Gruneisen temperature,  are shown in Table \ref{Table:I}. 
 Compared to the widely studied AHE induced by Gaussian disorder, we reveal significant difference in the phonon induced AHE: (a) While the disorder induced AHE is  insensitive to temperature, 
the phonon induced AHE depends on temperature significantly. Only at the high temperature limit $T\gg T_{BG}$, the unscreened DP induced AHE saturates to the AHE induced by Gaussian disorder. (b)While the side jump contribution due to phonon scatterings is finite as $T$ goes to zero,  both the phonon induced intrinsic skew scattering contribution and the coherent skew scattering contribution (which comes from the crossed diagrams) approach zero as $\sim T^2$ when the temperature $T$ goes to zero. This is in significant difference from the AHE induced by Gaussian disorder in a 2D massive Dirac metal, for which both 
the side jump and skew scattering contributions are finite as $\sim T^0$ in the whole temperature range. (c)Moreover, at low temperature, the phonon induced skew scattering contributions from the non-crossing and crossed diagrams cancel each other in the leading order of $\sim T^2$. The total phonon induced skew scattering contribution  is proportional to $\sim T^4$ at low temperature.

The side jump and intrinsic skew scattering contributions we obtained from the QFT for unscreened DP are consistent with the results from the semi-classical approach in~\cite{Niu2019, Xiao2019-1}.
We also studied the screening effect on the AH conductivity and found that the screening does not change the AH conductivity at the zero temperature limit but modifies the AH conductivities  at finite temperature. 

\begin{figure}[t]
	\includegraphics[width=8.5cm]{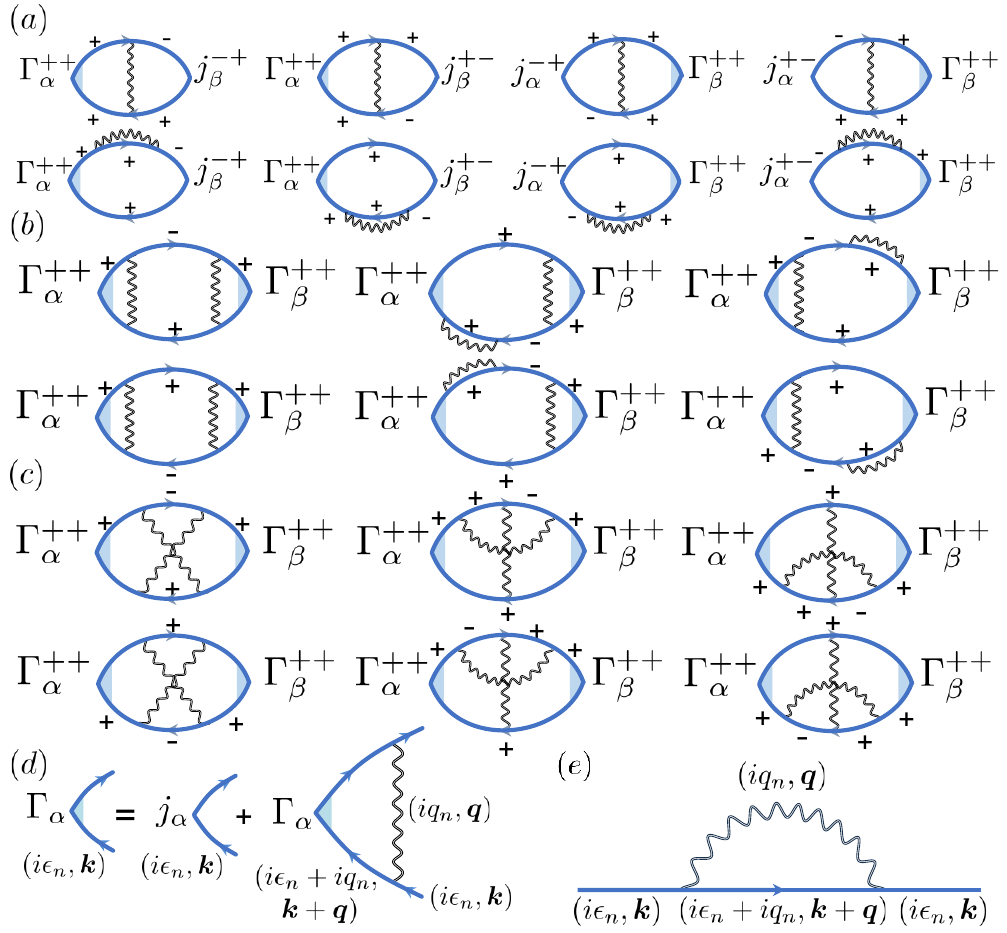}
		\caption{The Feynman diagrams of phonon induced (a) side jump, (b)non-crossing skew scattering and (c) coherent skew scattering contributions. The solid  and curvy lines represent the electron and phonon propagators respectively. The $+$ and $-$ label the propagators of the upper and lower band electrons respectively. (d)The depiction of the recursion equation of the renormalized current vertex. (e) The electron self-energy in the first Born approximation.}\label{fig:AHE_diagram}
\end{figure}

We start with a 2D massive Dirac model with 
\begin{equation}\label{eq:Dirac}
{\cal H}_0=v{\boldsymbol\sigma}\cdot {\mathbf k} +\Delta \sigma_z
\end{equation}
where $\boldsymbol \sigma=(\sigma_x, \sigma_y)$ is composed of Pauli matrices and $\Delta$ is the mass of the  Dirac fermion which breaks the time reversal symmetry of ${\cal H}_0$. The two energy bands of ${\cal H}_0$ are $\epsilon^{\pm}_{k}=\pm \sqrt{v^2 k^2+\Delta^2}$. The corresponding  eigenstates are $|u_{\mathbf k}^+\rangle=(\cos{\frac{\alpha}{2}}, \sin{\frac{\alpha}{2}} e^{i\theta_0})^T$ and $|u_{\mathbf k}^-\rangle=(\sin{\frac{\alpha}{2}}, -\cos{\frac{\alpha}{2}} e^{i\theta_0})^T$ where $\cos{\alpha}\equiv \Delta/|\epsilon^\pm_k|$ and $\theta_0$ is  the polar  angle of $\mathbf k$.

The electron-phonon interaction can be written  as 
\begin{equation}\label{eq:e_phonon}
{\cal H}_{ep}=\sum_{\mathbf{k, q}}\hat{\Psi}^\dag_{\mathbf{k+q}}\hat{M}(\mathbf q)\hat{\Psi}_{\mathbf k}(\hat{b}_{\mathbf q}+\hat{b}^\dag_{-\mathbf q}),
\end{equation}
where $\hat{\Psi}_{\mathbf k}$ and $\hat{b}_{\mathbf q}$ represent the electron and  phonon field respectively.
 For simplicity, in this work we focus on the AHE induced by acoustic DP for which the e-phonon interaction vertex can be written as a scalar as  
$ \hat{M}_{DP}(\mathbf q)=g_{\mathbf q}\equiv i q \xi_q g_D$, where $\xi_q\equiv\sqrt{ 1/2\rho \omega_{\mathbf q}},\ \omega_{\mathbf q}\approx s q$ is the phonon frequency, $\rho$ is the atomic mass density and $g_D$ is the DP strength. We ignore the screening effect on the DP at first and discuss its correction at the end of the calculation. 

We set the electron Fermi energy $\epsilon_F > \Delta$ as the largest energy scale in this work, i.e.,  $\epsilon_F \gg T, T_{BG}$. The phonon scattering induced AHE can be obtained by treating the phonons as  impurities excited by the  temperature.
We assume weak e-phonon interaction such that $\epsilon_F \tau\gg 1$ where $\tau$ is the mean lifetime of electrons. The AH conductivity is a sum of two contributions $\sigma_{xy}=\sigma^{\rm I}_{xy}+\sigma^{\rm II}_{xy}$ using the Kubo-Streda formula~\cite{Streda1982, Sinitsyn2007}. The quantity $\sigma_{xy}^{\rm I}$ may be considered as a contribution from electrons on the Fermi surface whereas $\sigma^{\rm II}$ is determined by all electron states under the Fermi surface. Since $\sigma^{\rm II}$ is insensitive to impurity scatterings at $\epsilon_F \tau\gg 1$~\cite{Sinitsyn2007} and vanishes for $\epsilon_F>\Delta$ for the 2D massive Dirac model, we focus on the study of $\sigma_{xy}^{\rm I}$. The contribution to   $\sigma_{xy}^{\rm I}$ from the non-crossing diagram in the spin basis can be written as~\cite{Sinitsyn2007}
\begin{equation}\label{eq:sigma_I_main}
  \sigma^{\rm I}_{xy}=-\sum_{\mathbf k}\int\frac{d\epsilon}{2\pi}\partial_{\epsilon} n_F(\epsilon) {\rm Tr}[\hat{\Gamma}_{x}G^R(\epsilon,\mathbf k)\hat{j}_{y}G^A(\epsilon,\mathbf k)], 
\end{equation}
where $G^{R/A}$ is the retarded/advanced electron Green's function (GF), and $\hat{j}_{y}$ and $\hat{\Gamma}_{x}$ are the bare and renormalized current vertex respectively.

The intrinsic contribution of $\sigma^{\rm I}$ is independent of phonon scattering and has been obtained in \cite{Sinitsyn2007} as $\sigma^{int}_{xy}=-e^2\Delta/4\pi \sqrt{\Delta^2+v^2 k^2_F}$. 
The phonon induced side jump and skew scattering contributions from the non-crossing diagram can be most easily separated  by expanding the trace in Eq.(\ref{eq:sigma_I_main}) in the eigenstate band basis~\cite{Zhang2023, Sinitsyn2007}. The results are depicted  in Fig.\ref{fig:AHE_diagram}(a) and (b). Besides, the skew scattering contribution can also come from the crossed diagrams shown in Fig.\ref{fig:AHE_diagram}(c). We first study  the side jump and skew scattering contribution  in the non-crossing limit, i.e., from Eq.(\ref{eq:sigma_I_main}) in the following and study the skew scattering contribution from the crossed diagram at the end.

The phonon propagator in the imaginary time formalism  is 
\begin{equation}
D_0(i q_n, \mathbf q)\equiv-\langle T_\tau u_{\mathbf q}u_{-\mathbf q}\rangle=\frac{2\omega_{\mathbf q}}{(iq_n)^2-\omega^2_{\mathbf q}},
\end{equation} 
where $u_{\mathbf q}\equiv b_{\mathbf q}+b^\dag_{-\mathbf q}$ and $iq_n=2n \pi  i/\beta, n\in Z$ is the phonon Matsubara frequency.
Here we do not include the phonon self-energy  due to e-phonon interaction explicitly because it only results in a renormalisation of the phonon velocity $s$. 
Therefore we only need to  assume the phonon velocity is the renormalized one. 

The leading order contribution to the AHE requires the electron GF in the first Born approximation $G(i\epsilon_n, \mathbf k)=[i\epsilon_n-{\cal H}_0-\Sigma(i\epsilon_n, \mathbf k)]^{-1}$~\cite{Sinitsyn2007, Zhang2023}, where the electron self-energy $\Sigma(i\epsilon_n, \mathbf k)$ due to e-phonon interaction is depicted in Fig.\ref{fig:AHE_diagram}(e). Since the phonon energy is much smaller than the electron Fermi energy $\epsilon_F$, we make the approximation that the electrons are bound to the Fermi surface both before and after the scattering with a phonon, i.e., the scatterings are quasi-elastic. The inclusion of the energy transfer during the scatterings only results in a correction smaller in the order of $T_{\rm BG}/\epsilon_F$ in the AH conductivity. With this approximation, we obtain the electron self-energy after analytic continuation to the real energy axis as (see Appendix A)
\begin{equation}
\Sigma^R(\epsilon, \mathbf k)\approx-\frac{i}{2}[a(1+\frac{\Delta}{\epsilon}\sigma_z)+v  \frac{b}{\epsilon}\boldsymbol \sigma\cdot \mathbf k],
\end{equation}
where
\begin{eqnarray}
 a&=&
    \frac{1}{4\pi} \frac{ g_D^2}{\rho v^2s^3}\frac{\epsilon}{k}
    \int_0^{k_B T_{BG}}  \Omega d \Omega \left(1-\frac{\Omega^2}{4 s^2 k^2}\right)^{-\frac{1}{2}} \nonumber\\
  &&\ \ \ \ \ \ \ \ \ \ [2n_B(\Omega)+1+n_F(\epsilon+\Omega)-n_F(\epsilon-\Omega)],  \label{eq:a_0} \\
b&=&
    \frac{1}{4\pi} \frac{ g_D^2 }{\rho v^2s^3}\frac{\epsilon}{k}
       \int_0^{k_B T_{BG}} \Omega d \Omega (1-\frac{\Omega^2}{2 s^2 k^2})(1-\frac{\Omega^2}{4 s^2 k^2})^{-\frac{1}{2}} 
       \nonumber\\
        &&\ \ \ \ \ \ \ \ \ \ [2n_B(\Omega)+1+n_F(\epsilon+\Omega)-n_F(\epsilon-\Omega)].\label{eq:b_0} 
\end{eqnarray}

The Feynman diagrams for the AH conductivity  include a vertex correction to the current operator by the e-phonon interaction, as shown in Fig.\ref{fig:AHE_diagram}(d). 
The leading order current vertex correction involves scatterings only within the upper electron band and the vertex correction due to such scatterings needs to be summed to infinite order~\cite{Sinitsyn2007}. Instead, for the vertex correction due to inter-band scatterings,  only the lowest order needs to be kept  in the calculation of the AH conductivity.
 As shown in Appendix B, the renormalized band-diagonal matrix element of the current vertex in the dc limit associated with the upper band, i.e., $\Gamma_{\alpha}^{++}(\epsilon, \epsilon;\mathbf k)\equiv \dirac {u_{\mathbf k}^+}{ \hat{\Gamma}_{\alpha}(\epsilon, \epsilon;\mathbf k)}  {u_{\mathbf k}^+}, \ \alpha=x, y$, satisfies the recursion equation
\begin{eqnarray}\label{eq:Re_Vertex}
&& \Gamma_{\alpha}^{++}(\epsilon, \epsilon; \mathbf k)= j^{++}_{\alpha}(\mathbf k)
 +\sum_{\mathbf q} \int d\xi |{g}_{\mathbf q }|^2   |\inner{u_{\mathbf k+\mathbf q}^+}{u_{\mathbf k}^+}|^2\nonumber\\
 &&\ \ \ \ \ G^{R+}(\xi,\mathbf  k+\mathbf q) G^{A+}(\xi,\mathbf  k+\mathbf q)\Gamma_{\alpha}^{++}(\xi,\xi; \mathbf  k+\mathbf q)  \nonumber\\
    &&\ \ \ \ \ \ \ \{ \delta(\xi-\epsilon-\omega_{\mathbf q})[n_B(\omega_{\mathbf q})+ n_F(\xi)] \nonumber\\
   &&\ \ \ \  \ \ \ \   +\ \delta(\xi-\epsilon+\omega_{\mathbf q})[n_B(\omega_{\mathbf q})+ 1-n_F(\xi)]\},
   \end{eqnarray}
where ${j}^{++}_{\alpha}(\mathbf k)=\dirac {u_{\mathbf k}^+}{ ev\sigma_{\alpha}}  {u_{\mathbf k}^+}=ev \frac{vk_\alpha}{\epsilon_k}$ is the bare current matrix element and 
\begin{equation}
G^{R/A, +}(\epsilon, \mathbf k)=\dirac{u_{\mathbf k}^+}{ G^{R/A}}{u_{\mathbf k}^+} = \frac{1 }{\epsilon -\epsilon_{k}^+ \pm \frac{i}{2\tau_k^+}}
\end{equation}
are the retarded and advanced GF of the upper band electrons. The upper band scattering rate is
\begin{equation}
1/\tau^+_k=(1+\frac{\Delta^2}{\epsilon \epsilon_k})a+\frac{v^2 k^2}{\epsilon\epsilon_k} b.
\end{equation}

\begin{table}
	\begin{center}
		\begin{tabular}{c|c|c}
		\hline
			\hline
			\ \ \ \ \ \ \ \ \ \ \ \ \  &\  \ $T\ll T_{BG} $ \ \ & \ \ $T\gg T_{BG}$  \\
			\hline
			$\sigma_{xy}^{\rm side}$ \   & $-\frac{e^2}{4\pi }\frac{\Delta}{\epsilon_F} (1-\frac{\Delta^2}{\epsilon_F^2}) \   $ & $ -\frac{e^2}{\pi }\frac{\Delta}{\epsilon_F} \frac{v^2k_F^2}{\epsilon_F ^2+3\Delta^2} $  \\
			$\sigma_{xy}^{\rm sk-nc}$  & \  $-\frac{\pi e^2}{2}\frac{\Delta}{\epsilon_F}(1-\frac{\Delta^2}{\epsilon^2_F})^2 \frac{T^2}{T^2_{BG}}$ \   &  \ $-\frac{3e^2}{4\pi }\frac{\Delta}{\epsilon_F} (\frac{v^2k_F^2}{\epsilon_F ^2+3\Delta^2})^2$ \  \\
			$\sigma_{xy}^{X+\Psi} $ &  $\frac{\pi e^2}{2}\frac{\Delta}{\epsilon_F}(1-\frac{\Delta^2}{\epsilon^2_F})^2 \frac{T^2}{T^2_{BG}}$ \   & $\frac{2e^2}{\pi}\frac{\epsilon_F \Delta (\epsilon^2_F - \Delta^2)}{(\epsilon^2_F+3\Delta^2)^2}$ \ \\
			a  & \  $\frac{\pi^2}{4}C (1+\frac{\pi^2}{4}t^2+\frac{3}{8}\pi^4 t^4)$   \ &  \ $\frac{\pi}{2} C (1/t-\frac{1}{12}t)$  \ \\
			b & \  $ \frac{\pi^2}{4}C(1-\frac{3\pi^2}{4} t^2 - \frac{5}{8}\pi^4 t^4)$   \  &  \  $\frac{\pi}{48} C/t^3$  \  \\
			c &  \ $\frac{\pi^2}{4}C(1-\frac{7\pi^2}{4}t^2+ \frac{19}{8}\pi^4 t^4)$  \ & \  $\frac{\pi}{4} C(1/t-\frac{1}{12}t)$  \  \\
			\hline
		\end{tabular}
	\end{center}	
	\caption{The expansion of the parameters $a, b, c$ with $(\epsilon, \mathbf k)$ on the Fermi surface  and the leading order AH conductivities  at the low and high temperature limit without screening, where $C\equiv\frac{1}{2\pi}\frac{g_D^2}{ \rho v^2 s^3 }\frac{\epsilon_F}{k_F}(k_B T )^2, t\equiv T/T_{BG}$  and we set $\hbar=1$.}
	\label{Table:I}
\end{table}

The recursion Eq.(\ref{eq:Re_Vertex}) is hard to solve exactly. But 
with the quasi-elastic scattering approximation, we can obtain the renormalized current vertex element $\Gamma_{\alpha}^{++}$ by an order by order iteration of Eq.(\ref{eq:Re_Vertex}) followed by a sum over all the orders, as shown in Appendix B. We  get the renormalized current vertex as
\begin{eqnarray}
&&\Gamma_{\alpha}^{++}(\epsilon, \epsilon; \mathbf k)=\gamma \frac{e v^2 k_\alpha}{\epsilon}, \\
&&\gamma=\frac{1}{1-\lambda}, \ \lambda= \frac{b+c+\frac{\Delta^2}{\epsilon^2}(b-c)}{a+b+\frac{\Delta^2}{\epsilon^2}(a-b)},
\end{eqnarray}
where $a$ and $b$ are given in Eq.(\ref{eq:a_0}) and (\ref{eq:b_0}) and  
\begin{eqnarray}
c&=&\frac{1}{4\pi}\frac{g_D^2}{\rho s^3 v^2}\frac{\epsilon}{k}
    \int_{0}^{k_B T_{BG}} \Omega d \Omega \  (1-\frac{\Omega^2}{2s^2 k^2})^2 (1-\frac{\Omega^2}{4s^2 k^2} )^{-\frac{1}{2}}   \nonumber\\
    &&\ \ \ \ \ \ \ \ \ [2n_B(\Omega)+1+n_F(\epsilon+\Omega)
     -n_F(\epsilon-\Omega))].\label{eq:c}
\end{eqnarray}

 It is interesting to note that  the  vertex correction factor $\gamma$ we obtained above from  Eq.(\ref{eq:Re_Vertex})  is equal to $\tau_k^{tr}/\tau_k^+$, where  $\tau_k^{tr}$ and $\tau_k^+$  are respectively  the  transport and mean lifetime  of the upper band electrons with phonon scatterings defined in Appendix B and~\cite{Niu2019}.

Since the AH conductivity $\sigma^{\rm I}_{xy}$ comes from electron scatterings on the Fermi surface, $\epsilon$ and $k$ in $a, b, c$ take the values $\epsilon_F$ and $k_F$ at the end of the calculation.
In the low and high  temperature limit, by expanding the parameters $a, b, c$ on the Fermi surface, as shown in Table \ref{Table:I},  we obtain
\begin{equation}
\gamma\approx \frac{1}{\pi^2}\left(\frac{T_{BG}}{T}\right)^2 \left[1+\frac{3\pi^2}{4}(1-2\frac{\Delta^2}{\epsilon^2_F})(\frac{T}{T_{BG}})^2\right]
\end{equation}
at $T\ll T_{BG}$ and 
\begin{equation}
\gamma\approx 2\frac{\epsilon_F^2+\Delta^2}{\epsilon_F^2+3\Delta^2}+\frac{1}{12}\left(\frac{T_{BG}}{T}\right)^2\frac{\epsilon_F^4+\Delta^4+6\epsilon_F^2\Delta^2}{(\epsilon_F^2+3\Delta^2)^2}
\end{equation}
at $T\gg T_{BG}$. It seems unusual that  $\gamma$ diverges as $\sim1/T^2$ when $T\to 0$. This is because both $1/\tau^+_k$ and $1/\tau_k^{tr}$ vanish as a power law of $T$ when $T\to 0$, but the transport scattering rate $1/\tau^{tr}_k \sim T^4$ vanishes faster than $1/\tau^+_k\sim T^2$. We will see later that this divergence of $\gamma$ at $T\to 0$ does not lead to the divergence of the AH conductivity at $T\to 0$.
At high temperature $T\gg T_{BG}$, $\gamma$ reduces to  $2\frac{\epsilon_F^2+\Delta^2}{\epsilon_F^2+3\Delta^2}$, which is the same as the current vertex renormalization factor due to Gaussian disorder~\cite{Sinitsyn2007}.

\begin{figure}
	\includegraphics[width=7.5cm]{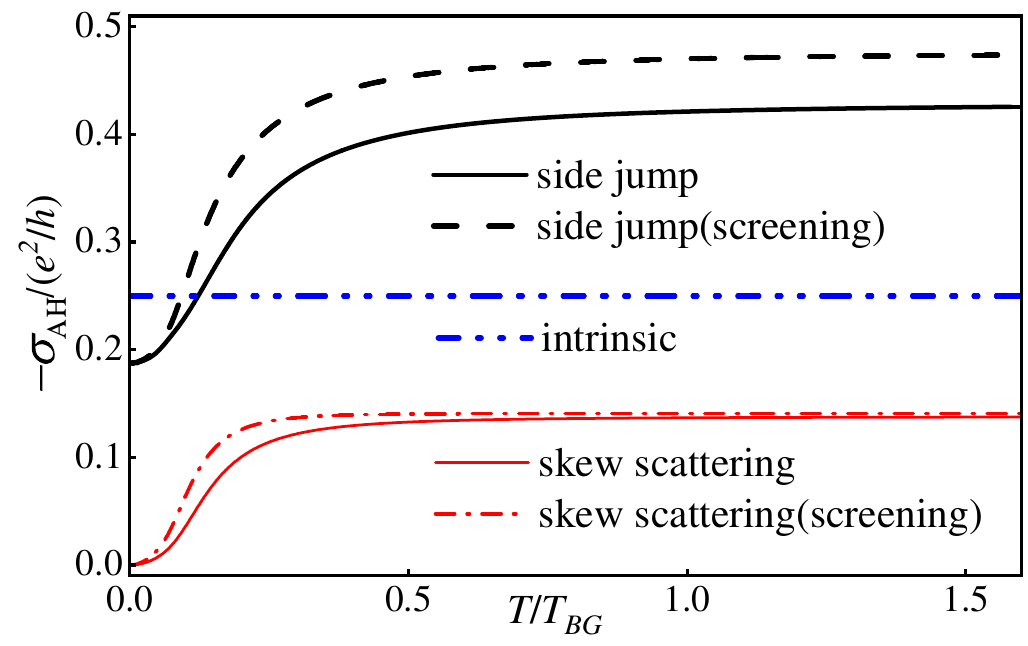}
			\caption{The intrinsic, side jump and skew scattering conductivities in the non-crossing limit as a function of $t=T/T_{BG}$ for $\Delta/\epsilon_F=1/2$ and $\alpha=2$. The dashed and solid lines represent the results with and without the screening effect respectively.}\label{fig:AH_conductivity}
\end{figure}

With the above ingredients, we can compute the side jump and intrinsic skew scattering conductivities due to phonon scatterings  depicted in Fig.\ref{fig:AHE_diagram}(a) and (b). 
After a lengthy calculation (see Appendix C), we obtain the two AH conductivities in the dc limit in a 2D massive Dirac metal as  
\begin{eqnarray}
\sigma_{xy}^{\rm side}&=& -\frac{e^2}{2\pi }\frac{\Delta}{\epsilon_F} (1-\frac{\Delta^2}{\epsilon_F^2})\frac{(a-b)}{a-c+\frac{\Delta^2}{\epsilon_F}(a+c-2b)}, \label{eq:side_jump_main}\\
\sigma^{\rm sk-nc}_{xy}&=& -\frac{e^2\Delta}{4\pi\epsilon_F}\left(1-\frac{\Delta^2}{\epsilon_F^2}\right)^2\frac{(a+c- 2b)(a-c)}{ [a-c+\frac{\Delta^2}{\epsilon_F}(a+c-2b)]^2}. \label{eq:skew_scattering_main}\nonumber\\
\end{eqnarray}
Defining $t\equiv T/T_{BG}$, the parameters $a, b, c$ with $\epsilon=\epsilon_F, k=k_F$ can be written as 
 \begin{eqnarray}
 a&=&C\int_0^{1/t}x dx (1- t^2 x^2)^{-\frac{1}{2}} F(x), \label{eq:tilde_a}\\
  b&=&C\int_0^{1/t} x dx 
   (1-2t^2 x^2)(1- t^2 x^2)^{-\frac{1}{2}}F(x), \\
   c&=&C\int_0^{1/t} x dx 
   (1-2t^2 x^2)^2(1- t^2 x^2)^{-\frac{1}{2}}F(x),
 \end{eqnarray}
where $F(x)\equiv \frac{1}{e^x-1}+\frac{1}{e^x+1}$ and $C\equiv \frac{1}{2\pi} \frac{ g_D^2}{\rho v^2s^3}\frac{\epsilon_F}{k_F}(k_B T)^2$.
The AH conductivities $\sigma_{xy}^{\rm side}$ and $\sigma^{\rm sk-nc}_{xy}$ then depend only on two parameters: $\Delta/\epsilon_F$ and $t$.  The numerical plots of $\sigma^{\rm side}_{xy}(t)$ and $\sigma^{\rm sk-nc}_{xy}(t)$  as a function of $t=T/T_{BG}$ for a given $\Delta/\epsilon_F=1/2$ are shown in Fig.\ref{fig:AH_conductivity}.

The analytical results of the phonon induced side jump and intrinsic skew scattering conductivities in the  limits $T\ll T_{BG}$ and $T\gg T_{BG}$ are shown in Table \ref{Table:I}. At $T\gg T_{BG}$, both the side jump and intrinsic skew scattering conductivities approach the values induced by Gaussian disorder in \cite{Sinitsyn2007}, indicating the saturation of phonon scatterings in this limit. At low temperature, however, the AH conductivity induced by deformation potential is significantly different from that induced by Gaussian disorder. 

At $T\ll T_{BG}$, we obtain the side jump contribution due to phonon scatterings as
\begin{equation}
\sigma^{\rm side}_{xy}\approx -\frac{e^2}{4\pi }\frac{\Delta}{\epsilon_F} (1-\frac{\Delta^2}{\epsilon_F^2})[1+2\pi^2(1-\frac{\Delta^2}{\epsilon^2_F})(\frac{T}{T_{BG}})^2 ]. 
\end{equation}
This result is consistent with the  side jump contribution at $T\to 0$ obtained from the SBE approach for phonon scatterings in \cite{Niu2019}, but  different from the result due to Gaussian disorder in \cite{Sinitsyn2007}.

For the phonon induced intrinsic skew scattering contribution,  the expansion of $a, b, c$ at $T\ll T_{BG}$ in Table \ref{Table:I}  gives
\begin{equation}
\sigma^{\rm sk-nc}_{xy}=-\frac{\pi e^2}{2}\frac{\Delta}{\epsilon_F}\left(1-\frac{\Delta^2}{\epsilon^2_F}\right)^2 \frac{T^2}{T^2_{BG}}+{\cal O}[(T/T_{BG})^4],
\end{equation}
i.e., the intrinsic skew scattering contribution approaches zero as $\sim(T/T_{BG})^2$ at $T\to 0$, as can be also seen from the numerical plot  in Fig.\ref{fig:AH_conductivity}. This is significantly different from the intrinsic skew scattering contribution  induced by Gaussian disorder with the noncrossing approximation, which is finite as $T\to 0$. We note that the vanishment of the phonon induced skew scattering contribution at $T\to 0$  from our quantum theory is consistent with the scaling analysis of the solution of the Boltzmann equation at low temperature for long range scalar impurity scatterings in \cite{Xiao2019-1}.

We next study the skew scattering contribution from the crossed or so-called $X$ and $\Psi$ diagrams in Fig.\ref{fig:AHE_diagram}c due to phonon scatterings. The importance of such diagrams to the skew scattering contribution was demonstrated in recent years for systems with Gaussian disorder~\cite{Ado2015, Ado2016, Ado2017, Levchenko2017, Chen2023}.
It is difficult to get a quantitative result of this contribution for phonon scatterings in the whole temperature regime due to the complexity of the calculation.
But we are able to obtain the contribution from the crossed diagrams due to phonon scatterings in both the low and high temperature limits, as shown in Table \ref{Table:I}. 
At high temperature $T\gg T_{BG}$, this contribution reduces to that of the crossed diagrams with Gaussian disorder, which has been studied in Ref.\cite{Ado2015}. In the low temperature limit $T\ll T_{BG}$, the e-phonon interaction is dominated by the small phonon momentum scatterings. By expansion in terms of the phonon momentum in the calculation and keeping only the leading order contribution, as shown in Appendix C, we found that the skew scattering contribution from the crossed diagrams is exactly opposite to the skew scattering contribution from the non-crossing diagrams in the leading order, which is proportional to $ T^2$, as shown in Table \ref{Table:I}. The total skew scattering contribution at low temperature is proportional to $\sim T^4$.

The above calculation ignored the e-e interaction~\cite{Principi2024}. For simplicity, we only consider the screening effects. To take into account this effect, we add the Thomas-Fermi (TF) screening factor to the deformation potential by replacing $g_D$ with $g_D\frac{q}{q+q_{TF}}$, where $q_{TF}\sim \alpha \epsilon_F/v$ is the TF wave vector and $\alpha=e^2/\hbar v$ is the fine structure constant~\cite{Chen2012, Oppen2010}.  The AH conductivities for non-crossing diagrams including the screening effect are plotted in Fig.\ref{fig:AH_conductivity} (for which we set $\alpha=2$ as for graphene). One can see that the inclusion of screening does not change the AH conductivites at $T\to 0$. Particularly, for the intrinsic skew scattering contribution, $\sigma^{\rm sk-nc}_{xy}$ still vanishes as $\sim T^2/T^2_{BG}$ at $T\to 0$ but with a modified coefficient as $\tilde{\sigma}^{\rm sk-nc}_{xy}\approx \frac{17}{8}\sigma^{\rm sk-nc}_{xy}$. The same happens to the coherent skew scattering contribution at low temperature.
At finite temperature, the screening effect modifies the AH conductivities and their limiting values at $T\gg T_{BG}$
 depend on $\alpha$. More detailed discussion of the screening effect is shown in Appendix D.

The temperature dependence of the phonon induced AH conductivity has been pointed out and analyzed in previous works with the semi-classical approach \cite{Niu2019, Xiao2019-1, Xiao2019}. 
This is in contrast to the AHE due to Gaussian disorder for which the AH conductivity is independent of the temperature. The reason is because the AH conductivity depends on the scattering range~\cite{Ado2017}, which depends on the temperature $T$ for phonon scatterings~\cite{Niu2019, Xiao2019} but independent of $T$ for Gaussian disorder. At $T\gg T_{BG}$, the phonons participating in the scatterings saturate and the momentum transfer during the scatterings is randomly distributed from $0$ to $2k_F$. The average momentum transfer, or the scattering range then approaches that for Gaussian disorder in \cite{Sinitsyn2007}, so does the AH conductivity. 
The quantum approach in this work provides a microscopic and quantitative description of the crossover from the low to high temperature of the AH conductivity due to phonon scatterings.


The phonon induced AHE and its temperature dependence we discussed above may be observed in clean 2D Dirac metals with TRS breaking, such as $\rm Fe_3Sn_2$, which is a quasi-2D ferromagnetic Dirac metal~\cite{Fu2021},  or graphene with spin-orbit interaction and TRS breaking~\cite{MacDonald2006}. The spin-orbit interaction results in a gap or finite mass in the graphene. The TRS breaking avoids the cancellation of the AH conductivities from the two valleys and may be achieved by spin polarization of the graphene through optical orientation~\cite{Belinicher1980}, or ferromagnetic contacts~\cite{MacDonald2006}.

{\it Acknowledgement.} We thank Cong Xiao for very helpful discussion. This work is supported by the NNSF of China under Grant No.11974166 and the Department of Science and Technology of Jiangsu Province under Grant No. BK20231398.

\begin{widetext}

\section{Appendix A: Electron Self-energy and Green's function in the first Born Approximation}

\subsection{Electron Self-energy}

We show the detailed calculation of the electron self-energy due to the e-phonon interaction and the electron GF in the first Born approximation in this appendix.

 The bare electron Matsubara GF of the 2D massive Dirac model is 
 \begin{equation}
 G_0(i\epsilon_n,\mathbf k)=\frac{1}{(i\epsilon_n)^2-\epsilon_{k}^2}(i\epsilon_n\sigma_0 + \Delta \sigma_z+ v\mathbf k \cdot \boldsymbol \sigma),
 \end{equation}
 where $i\epsilon_n=(2n+1)\pi i/\beta$  is the electron Matsubara frequency with $\beta=1/k_BT, n\in Z$.

%



For the e-phonon interaction given in Eq.(\ref{eq:e_phonon}), the electron self-energy in the first Born approximation depicted in Fig.\ref{fig:self_energy}(a) can be expressed as 
\begin{equation}
 \Sigma(i\epsilon_n,\mathbf k)=-\frac{1}{\beta}\sum_{iq_n}\sum_{\mathbf q}|g_{\mathbf q}|^2  D_0 (iq_n, \mathbf q)G_0(i\epsilon_n+iq_n,\mathbf k+\mathbf q), 
\end{equation}
where $D_0(iq_n, \mathbf q)$ is the bare phonon propagator given in the main text.  The sum over the phonon Matsubara frequency $iq_n$ may be obtained by performing the following integral over the contour in Fig.\ref{fig:self_energy}b:
\begin{eqnarray}\label{eq:self_energy_contour}
 &&\int_{\cal C}\frac{dz}{2\pi i}n_{B}(z)D_0(z, \mathbf q)G_0(z+i\epsilon_n, \mathbf k+\mathbf q) \nonumber\\
   =&&\int_{-\infty}^\infty \frac{d \xi}{2\pi i}n_B(\xi -i\epsilon_n)D_0(\xi-i\epsilon_n, \mathbf q)[G_0(\xi+i0^+,\mathbf k+\mathbf q)-G_0(\xi-i0^+, \mathbf k+\mathbf q)]  \nonumber\\
  =&&\frac{1}{\beta}\sum_{iq_n}D_0(iq_n, \mathbf q)G_0(iq_n+i\epsilon_n, \mathbf k+\mathbf q)
 +   \sum_{z_j=\pm \omega_q}{\rm Res} [D_0(z=z_j, \mathbf q)]G_0(z_j+i\epsilon_n, \mathbf k+\mathbf q)n_B(z_j),
\end{eqnarray}
where $n_B(z)$ is the Bose-Einstein distribution function and ${\rm Res} [D_0(z=z_j, \mathbf q)]$ is the residue of $D_0(z, \mathbf q)$ at $z=z_j$.

\begin{figure}
	\includegraphics[width=17cm]{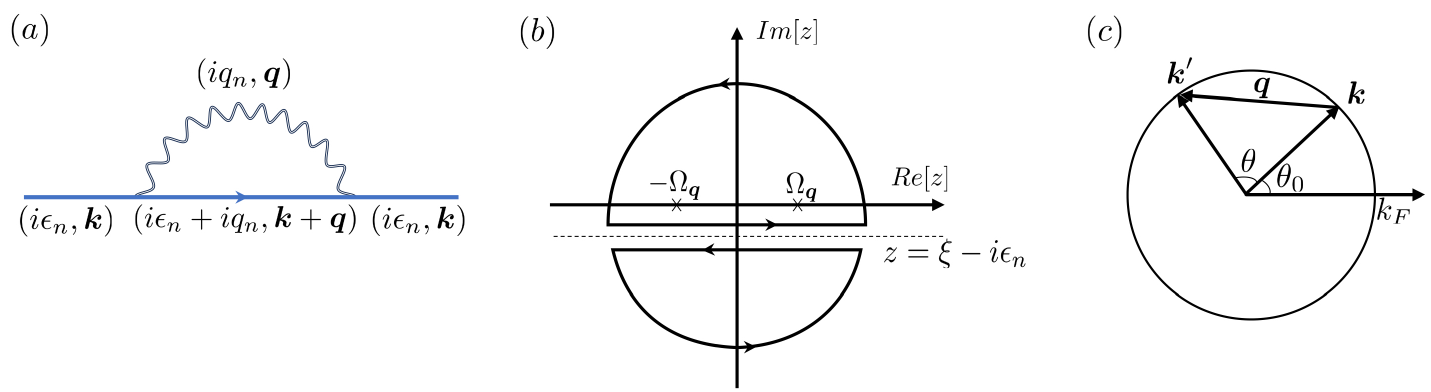}
			\caption{(a)Feynman diagram of the electron self-energy due to e-phonon interaction. The solid and curvy lines represent the electron and phonon propagators respectively. (b)The integration contour for the summation of the phonon Matsubara frequency of the self-energy. (c)Quasi-elastic scattering of an electron off a phonon near the Fermi surface. Here $\mathbf k$ and $\mathbf k'$ are the initial and final momentum of the electron and $\mathbf q$ is the momentum of the phonon.}\label{fig:self_energy}
\end{figure}
 
We assume that the e-phonon interaction is weak so the real part of the self energy is much smaller than the Fermi energy and we can ignore it. We then only need to compute the imaginary part of the electron self-energy. From Eq.(\ref{eq:self_energy_contour}), we obtain the self-energy  (i.e., imaginary part) after the sum over                                                            
the Matsubara frequency $iq_n$ and the analytic continuation $i\epsilon_n\to \epsilon+i 0^+$ to real energy axis as
\begin{eqnarray}
\Sigma^R(\epsilon,\mathbf k)
    &=&(-i\pi)\sum_{\mathbf q}|g_{\mathbf q}|^2 \int_{-\infty}^\infty \frac{d\xi}{2\xi} \delta(\xi- \epsilon_{\mathbf k+\mathbf q}) [\xi\sigma_0 + \Delta \sigma_z+ v(\mathbf k+\mathbf q) \cdot \boldsymbol \sigma]\\
    &&\times 
    [ \delta(\xi-\epsilon-\omega_q)(n_B(\omega_q)+n_F(\xi))- \delta(\xi-\epsilon+\omega_q)(n_B(-\omega_q)+n_F(\xi))],
\end{eqnarray}
where $n_F(\xi)$ is the Fermi-Dirac distribution function and we have used  $n_B(\xi -i\epsilon_n)=-n_F(\xi)$.

For an electron with momentum $\mathbf k$ on the Fermi surface, and the phonon energy $\omega_q$ much smaller than the Fermi energy $\epsilon_F$, the electron after scattering with a phonon is still very close to the Fermi surface so the maximum momentum (energy) of the phonon participating in the scatterings is about $2k_F\ (2sk_F)$. The sum over the phonon momentum $\mathbf q$ in the self-energy may be converted to the integral over $\mathbf k'=\mathbf k+\mathbf q$ as
\begin{eqnarray}                     
\Sigma^R(\epsilon,\mathbf k) 
    &=&\frac{(-i\pi)}{(2\pi)^2}\frac{g_D^2 }{2 \rho s^2}\int_{-\infty}^\infty \frac{d\xi}{2\xi} \int_{0}^{k_BT_{BG}} \Omega d \Omega  
     \nonumber\\
    &&
    \left[ \delta(\xi-\epsilon-\Omega)(n_B(\Omega)+n_F(\xi))+ \delta(\xi-\epsilon+\Omega)(n_B(\Omega)+1-n_F(\xi))\right]\nonumber\\
    &&\int _{0}^{2\pi}d\theta \delta(\Omega-\omega_q)\int k'd k' \delta(\xi- \epsilon_{\mathbf k'})\times 
    \left( \begin{array}{ll}
       \xi+\Delta  &  vk'e^{-i\theta}e^{-i\theta_0} \\\\
       vk'e^{i\theta}e^{i\theta_0}  &\ \ \ \  \xi-\Delta
    \end{array} 
    \right),
\end{eqnarray}
where $\theta_0$ is the polar angle of $\mathbf k$,   $\theta$ is the angle between $\mathbf k$ and $\mathbf k'$ as shown in Fig.\ref{fig:self_energy}(c) and $\omega_q\approx 2s k \sin\frac{\theta}{2}$.  
We have  introduced an  integration over $d\Omega$ through the factor $\delta(\Omega-\omega_q)$ in the above equation. This procedure converts the integration over the angle $d\theta$ to the integration over $d\Omega$ through the relationship $\omega_q\approx 2sk\sin\frac{\theta}{2}$.
Since $\epsilon_{ k'}=\sqrt{v^2 k'^2+\Delta^2}$, $k' dk'=\frac{\epsilon_{  k'}}{v^2}d\epsilon_{  k'}$, the integration over $k'$ can be converted to $\epsilon_{ k'}$. 
After the integration over $\epsilon_{  k'}$ and $\theta$, we get 
\begin{eqnarray}
\Sigma^R(\epsilon,\mathbf k) & =& - \frac{i}{4\pi}\frac{ g_D^2 }{2\rho v^2s^4} \int_{0}^{k_B T_{BG}} \Omega^2 d \Omega   \frac{1}{kp}[1-(\frac{k^2+p^2-(\Omega/s)^2}{2kp})^2]^{-1/2}\nonumber\\
&&\int_{-\infty}^\infty d\xi  [ \delta(\xi-\epsilon-\Omega)(n_B(\Omega)+n_F(\xi))
+ \delta(\xi-\epsilon+\Omega)(n_B(\Omega)+1-n_F(\xi))]\nonumber\\
&&
  \left(
    \begin{array}{ll}
       \xi+\Delta  &  vp e^{-i\theta_0}\frac{k^2+p^2-(\Omega/s)^2}{2kp}\\\\
       vpe^{i\theta_0} \frac{k^2+p^2-(\Omega/s)^2}{2kp} &\ \ \ \ \ \ \ \  \xi-\Delta
    \end{array} 
    \right),
\end{eqnarray}
where $p=\frac{\sqrt{\xi^2-\Delta^2}}{v}$.

We can write the above self-energy as 
\begin{equation}
 \Sigma^R(\epsilon,\mathbf k)=-\frac{i}{2}  
 \left(
 \begin{array}{ll}
    a+ \frac{\Delta}{\epsilon}\tilde{a} & v(k_x-ik_y)\frac{b}{\epsilon} \\
    v(k_x+ik_y)\frac{b}{\epsilon}  & \ \ \ a- \frac{\Delta}{\epsilon}\tilde{a}
 \end{array}\right),
 \end{equation}
where the parameters $a, \tilde{a}, b$ are 
\begin{eqnarray}\label{eq:a}
    a(\epsilon, k)&=
    &\frac{1}{4\pi} \frac{ g_D^2 }{\rho v^2s^4}\frac{1}{k}
    \int_{-\infty}^\infty d\xi  
    \int_0^{k_B T_{BG}}  \Omega^2 d \Omega\frac{1}{p}[1-(\frac{k^2+p^2-(\Omega/s)^2}{2kp})^2]^{-1/2} \nonumber\\
    &&\times \xi[ \delta(\xi-\epsilon-\Omega)(n_B(\Omega)+n_F(\xi))
+ \delta(\xi-\epsilon+\Omega)(n_B(\Omega)+1-n_F(\xi))],
\end{eqnarray}
\begin{eqnarray}\label{eq:tilde_a}
    \tilde{a}(\epsilon, k)&=
    &\frac{\epsilon}{4\pi} \frac{ g_D^2 }{\rho v^2s^4}\frac{1}{k}
    \int_{-\infty}^\infty d\xi  
    \int_0^{k_B T_{BG}}  \Omega^2 d \Omega \frac{1}{p}[1-(\frac{k^2+p^2-(\Omega/s)^2}{2kp})^2]^{-1/2} \nonumber\\
    &&\times [ \delta(\xi-\epsilon-\Omega)(n_B(\Omega)+n_F(\xi))
+ \delta(\xi-\epsilon+\Omega)(n_B(\Omega)+1-n_F(\xi))],
\end{eqnarray}
\begin{eqnarray}\label{eq:b}
   b(\epsilon,  k)&=
    &\frac{\epsilon}{4\pi} \frac{ g_D^2 }{\rho v^2s^4}\frac{1}{k^2}
    \int_{-\infty}^\infty d\xi  
    \int_0^{k_B T_{BG}} \Omega^2 d \Omega \frac{k^2+p^2-(\Omega/s)^2}{2kp}[1-(\frac{k^2+p^2-(\Omega/s)^2}{2kp})^2]^{-1/2} \nonumber\\
    && \times [ \delta(\xi-\epsilon-\Omega)(n_B(\Omega)+n_F(\xi))
+ \delta(\xi-\epsilon+\Omega)(n_B(\Omega)+1-n_F(\xi))].
\end{eqnarray}

In this work, we are interested in the AH conductivity  $\sigma^{\rm I}$ which comes from the contribution of electrons on the Fermi surface. For the reason, $(\epsilon, \mathbf k)$ is bound to the Fermi surface. The electron energy after scattering with a phonon is $\xi=\epsilon\pm \omega_q$. Since $T_{BG}\ll \epsilon_F$  in our setting, the phonon scattering is quasi-elastic, i.e., $\xi\approx \epsilon=\epsilon_F$ in Eq.(\ref{eq:a})-(\ref{eq:b}) and $a\approx \tilde{a}$. 
The self energy can then be written as        
\begin{equation}
\Sigma^R(\epsilon,\mathbf k) \approx-\frac{i}{2}[a(1+\frac{\Delta}{\epsilon}\sigma_z)+v\frac{b}{\epsilon}\boldsymbol \sigma\cdot \mathbf k)]
\end{equation}                                                                  
as in the main text, where $a, b$ can be simplified as 
\begin{eqnarray}\label{eq:simplified_a}
    a(\epsilon,  k)&=
    &\frac{1}{4\pi} \frac{ g_D^2}{\rho v^2s^3}\frac{\epsilon}{k}
    \int_0^{k_B T_{BG}}  \Omega d \Omega \left(1-\frac{\Omega^2}{4 s^2 k^2}\right)^{-\frac{1}{2}} 
    [2n_B(\Omega)+1+n_F(\epsilon+\Omega)-n_F(\epsilon-\Omega)],
\end{eqnarray}
\begin{eqnarray}\label{eq:simplified_b}
    b(\epsilon, k)&=
    &\frac{1}{4\pi} \frac{ g_D^2 }{\rho v^2s^3}\frac{\epsilon}{k}
       \int_0^{k_B T_{BG}} \Omega d \Omega \left(1-\frac{\Omega^2}{2 s^2 k^2}\right)\left(1-\frac{\Omega^2}{4 s^2 k^2}\right)^{-\frac{1}{2}} 
         [2n_B(\Omega)+1+n_F(\epsilon+\Omega)-n_F(\epsilon-\Omega)].
\end{eqnarray}

At the end of the calculation of $\sigma^{\rm I}_{xy}$, we set $k=k_F, \epsilon=\epsilon_k=\epsilon_F$. At $T\ll T_{BG}$ and $T\gg T_{BG}$, we can expand the integrand in Eq.(\ref{eq:simplified_a}) and (\ref{eq:simplified_b}) and get the analytic results of $a$ and $b$ in the two limits as shown in Table I in the main text.

\subsection{Electron Green's function in the first Born approximation}

The electron GF in the first Born approximation is 
\begin{eqnarray}\label{eq:GF_BA}
G^R(\epsilon,\mathbf k)
    &=&[G_0^{-1}(\epsilon,\mathbf k)-\Sigma^R(\epsilon,\mathbf k)]^{-1} \nonumber\\
    &=& \left( 
\begin{array}{ll}
   \epsilon-\Delta+\frac{i}{2}a(1+\frac{\Delta}{\epsilon})  & -v(k_x-ik_y)(1- \frac{i}{2}\frac{b}{\epsilon}) \\\\
  -v(k_x+ik_y)(1- \frac{i}{2}\frac{b}{\epsilon})   & \ \ \ \epsilon+\Delta+\frac{i}{2}a(1-\frac{\Delta}{\epsilon})
\end{array}
    \right)^{-1} \nonumber\\
    &=&\frac{1}{\epsilon-\epsilon_k^++\frac{i}{2\tau_k^+}}\frac{1}{\epsilon-\epsilon_k^{-}+\frac{i}{2\tau_k^-}}
    \left[(1+\frac{i}{2\epsilon}a)\epsilon+ (1-\frac{i}{2\epsilon}a)\Delta\sigma_z+ (1-\frac{i}{2\epsilon}b)v\mathbf k\cdot \boldsymbol \sigma \right],
\end{eqnarray}
where $\epsilon_k^{\pm}$ are the two energy bands of ${\cal H}_0$ and 
\begin{equation}\label{eq:tau^pm}
 1/\tau^{\pm}_k=a\pm \frac{v^2k^2 b+\Delta^2a}{\epsilon\epsilon_k}.
 \end{equation}

The above GF can be written in the band basis as 
\begin{equation}\label{eq:GF_band}
 G^{R}(\epsilon,\mathbf k)=\frac{\ket{u^+_{\mathbf k}}\bra{u^+_{\mathbf k}}  }{\epsilon -\epsilon_{k}^++  \frac{i}{2\tau_k^+}}+ \frac{\ket{u_{\mathbf k}^-}\bra{u_{\mathbf k}^-}  }{\epsilon -\epsilon_{k}^- + \frac{i}{2\tau_k^-
    }},
\end{equation}
where $\ket{\mathbf k,\pm}$ are the two eigenvectors of ${\cal H}_0$.

\section{Appendix B: Vertex correction}

\subsection{Recursion equation of the renormalized current vertex}

\begin{figure}
	\includegraphics[width=17cm]{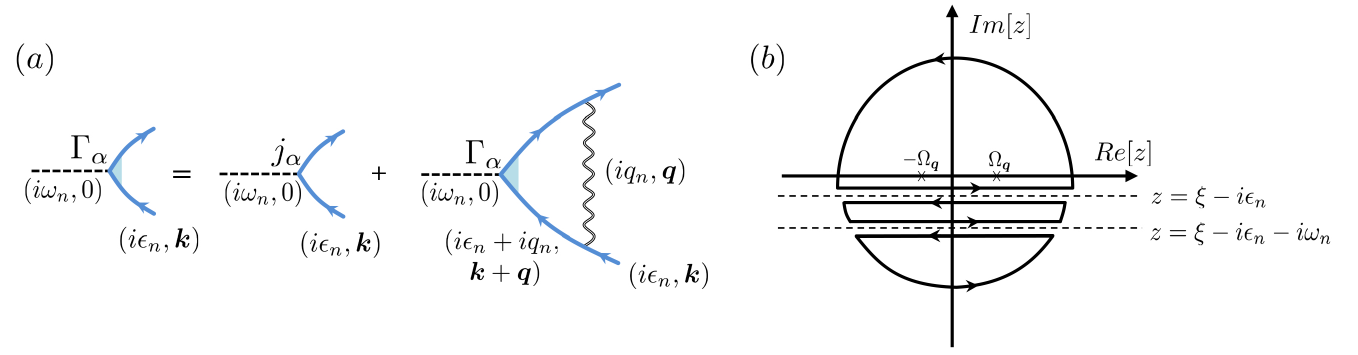}
			\caption{(a)Feynman diagram of the recursion equation of the renormalized current vertex. The solid and curvy lines represent the electron and phonon propagators respectively. (b)The integration contour for the summation of the phonon Matsubara frequency in the recursion equation of the renormalized current vertex.}\label{fig:Vertex_diagram}
\end{figure}

The renormalized current vertex is shown in  Fig.\ref{fig:Vertex_diagram}(a) and satisfies the recursion equation
\begin{eqnarray}\label{eq:vertex_correction}
  \hat{\Gamma}_{\alpha}( i\epsilon_n+i\omega_n, i\epsilon_n;\mathbf k)=\hat{j}_{\alpha} -\frac{1}{\beta}\sum_{i q_n, \mathbf{q}}   
 |g_{\mathbf q}|^2D_0(i q_n, \mathbf q)G(i \epsilon'_n+i \omega_n, \mathbf k')\hat{\Gamma}_{\alpha}(i\epsilon'_n+i\omega_n, i\epsilon'_n; \mathbf k')G(i\epsilon'_n, \mathbf k') 
  \end{eqnarray}
where $i\epsilon'_n\equiv i\epsilon_n+i q_n, \mathbf k'\equiv \mathbf k+\mathbf q$, $\hat{j}_\alpha=ev\sigma_\alpha$ is the bare current vertex, $i\omega_n$ is the external frequency of the vertex and  we have set the external momentum of the vertex to be zero.

We may express the current vertex in the Pauli matrix basis as 
\begin{equation}
 \hat{\Gamma}_{\alpha}(i\epsilon_n+i\omega_n, i\epsilon_n;\mathbf k)=ev \Lambda_{\alpha \beta}( i\epsilon_n+i\omega_n, i\epsilon_n; \mathbf k)\sigma_{\beta}, 
 \end{equation}
where $\alpha, \beta=0, x, y, z$ and the sum over repeated indices is implied in the whole text.

The recursion Eq.(\ref{eq:vertex_correction}) then becomes
\begin{equation}\label{eq:vertex_recursion}
  \Lambda_{\alpha \gamma}( i\epsilon_n+i\omega_n, i\epsilon_n;\mathbf k)
    =\delta_{\alpha\gamma}-\frac{1}{\beta}\sum_{iq_n}\sum_{\mathbf q}D_0(iq_n, \mathbf q)\Lambda_{\alpha\beta}( i\epsilon'_n+i\omega_n,i\epsilon'_n;\mathbf k+\mathbf q){\cal I}_{\beta\gamma}(i\epsilon'_n+i\omega_n, i\epsilon'_n;\mathbf k+\mathbf q),
\end{equation}
where
\begin{equation}
    {\cal I}_{\beta\gamma}(i\epsilon'_n+i\omega_n, i\epsilon'_n;\mathbf k+ \mathbf q)=\frac{1}{2}
    \Tr[\sigma_{\beta} G(iq_n+i\epsilon_n+i\omega_n, \mathbf k+\mathbf q)\sigma_{\gamma}G(iq_n+i\epsilon_n, \mathbf k+\mathbf q)]
\end{equation}
is the polarization operator.

The sum over the Matsubara frequency in Eq.(\ref{eq:vertex_recursion}) may be done by performing the contour integral in Fig.\ref{fig:Vertex_diagram}(b). Denoting
\begin{equation}   
Q(i\epsilon_n+i\omega_n, i\epsilon_n)\equiv-\frac{1}{\beta}\sum_{iq_n}D_0(iq_n, \mathbf q)\Lambda_{\alpha\beta}(iq_n+i\omega_n+i \epsilon_n, iq_n+i\epsilon_n; \mathbf k+\mathbf q){\cal I}_{\beta\gamma}(iq_n+i\epsilon_n+i\omega_n, iq_n+i\epsilon_n; \mathbf k+\mathbf q)\\
\end{equation}
and 
\begin{equation}
S(i\epsilon_n+i\omega_n, i\epsilon_n)\equiv\int_{\cal C}\frac{dz}{2\pi i}n_B(z)D_0(z, \mathbf q)\Lambda_{\alpha\beta}(z+i\epsilon_n+i\omega_n, z+i\epsilon_n;\mathbf k+\mathbf q)
    {\cal I}_{\beta\gamma}(z+i\epsilon_n+i\omega_n, z+i \epsilon_n; \mathbf k+\mathbf q),
\end{equation}
where ${\cal C}$ is the integration contour in Fig.\ref{fig:Vertex_diagram}(b), we get 
\begin{eqnarray}\label{eq:S}
&&S( i \epsilon_n+i\omega_n, i \epsilon_n)=-Q( i\epsilon_n+i\omega_n, i\epsilon_n)\nonumber\\
  &&  +\sum_{z_j=\pm \omega_q}Res [D_0(z=z_j, \mathbf q)]\Lambda_{\alpha\beta}(z_j+i\epsilon_n+i\omega_n, z_j+i\epsilon_n; \mathbf k+\mathbf q) {\cal I}_{\beta\gamma}(z_j+i\epsilon_n+i\omega_n, z_j+i\epsilon_n; \mathbf k+\mathbf q)n_B(z_j).
\end{eqnarray}

The contour integral $S$ on the circle vanishes and the integral becomes
\begin{eqnarray}
&&S( i\epsilon_n+i\omega_n, i\epsilon_n)
    =\int_{-\infty}^{\infty}\frac{d\xi}{2\pi i}n_B(\xi-i\epsilon_n)D_0(\xi-i\epsilon_n) \nonumber\\
    &&\ \ \ \ \ \ \ \ \ \ \  [\Lambda_{\alpha\beta}(\xi+i\omega_n, \xi+i0^+){\cal I}_{\beta\gamma}(\xi+i\omega_n, \xi+i0^+)-
    \Lambda_{\alpha\beta}(\xi+i\omega_n, \xi-i0^+){\cal I}_{\beta\gamma}(\xi+i\omega_n, \xi-i0^+)] \nonumber\\
    &&\ \ \   \ \ \ \ \ \ \ \ \  \ \ \ \ \ \ \ \ \ +\int_{-\infty}^{\infty} \frac{d\xi}{2\pi i}n_B(\xi-i\epsilon_n-i\omega_n)D_0(\xi-i\epsilon_n-i\omega_n) \nonumber\\
    &&\ \ \ \ \ \ \ \ \ \ \  [\Lambda_{\alpha\beta}(\xi+i0^+, \xi-i\omega_n){\cal I}_{\beta\gamma}(\xi+i0^+, \xi-i\omega_n)-
    \Lambda_{\alpha\beta}(\xi-i0^+, \xi-i\omega_n){\cal I}_{\beta\gamma}(\xi-i0^+,\xi-i\omega_n)].
\end{eqnarray}
For brevity we have dropped the momentum appeared in Eq.(\ref{eq:S})  in the above equation.

The dominant vertex correction comes from the $G^RG^A$ or $G^AG^R$ term in the polarization operator ${\cal I}$. For the reason, we perform the analytic continuation $i\epsilon_n \rightarrow \epsilon -i0^+, i\omega_n\rightarrow  \omega +i0^+$ and get
\begin{eqnarray}
 &&\ \ \ \ \ \ S^{RA}(\epsilon+\omega, \epsilon)=S( i\epsilon_n+i\omega_n\rightarrow \epsilon+\omega +i0^+, i\epsilon_n \rightarrow \epsilon -i0^+)
     \nonumber\\
    &&\ \  \   =-\int_{-\infty}^{\infty} \frac{d\xi}{2\pi i}n_F(\xi)D^R_0(\xi-\epsilon)[\Lambda^{RR}_{\alpha\beta}(\xi+\omega, \xi){\cal I}_{\beta\gamma}^{RR}(\xi+\omega, \xi)-
    \Lambda_{\alpha\beta}^{RA}(\xi+\omega, \xi){\cal I}_{\beta\gamma}^{RA}(\xi+\omega, \xi)] \nonumber\\
    &&\ \ \ \ \ \  -\int_{-\infty}^{\infty} \frac{d\xi}{2\pi i}n_F(\xi))D^A_0(\xi-\epsilon-\omega)
    [\Lambda_{\alpha\beta}^{RA}(\xi, \xi-\omega){\cal I}_{\beta\gamma}^{RA}(\xi, \xi-\omega)-
    \Lambda_{\alpha\beta}^{AA}(\xi, \xi-\omega){\cal I}^{AA}_{\beta\gamma}(\xi, \xi-\omega)] \nonumber\\
    &&\ \ \   \approx \int_{-\infty}^{\infty} \frac{d\xi}{2\pi i} \left[n_F(\xi)D_0^R(\xi-\epsilon)
    \Lambda_{\alpha\beta}^{RA}(\xi+\omega, \xi){\cal I}_{\beta\gamma}^{RA}(\xi+\omega, \xi) 
    - n_F(\xi)D_0^A(\xi-\epsilon-\omega)
    \Lambda_{\alpha\beta}^{RA}(\xi, \xi-\omega){\cal I}_{\beta\gamma}^{RA}(\xi, \xi-\omega)\right]. \nonumber\\
\end{eqnarray}
In the last equation, we dropped the ${\cal I}^{RR}$ and ${\cal I}^{AA}$ terms because they are small compared to the ${\cal I}^{RA}$ terms.

We are interested in the dc AH conductivity so we take the dc limit $\omega\to 0$ at the end and get 
\begin{equation}
  S^{RA}(\epsilon, \epsilon)
    = \int_{-\infty}^{\infty}\frac{d\xi}{2\pi i}n_F(\xi)[D_0^R(\xi-\epsilon)-D_0^A(\xi-\epsilon)]
    \Lambda_{\alpha\beta}^{RA}(\xi,\xi){\cal I}_{\beta\gamma}^{RA}(\xi,\xi).
\end{equation}

Since 
\begin{equation}
  D_0^R(\xi-\epsilon)-D_0^A(\xi-\epsilon)=-2i \pi [ \delta(\xi-\epsilon-\omega_q)-\delta(\xi-\epsilon+\omega_q)],
\end{equation}
we get 
\begin{equation}\label{eq:S_real}
S^{RA}(\epsilon, \epsilon)
    = -\int_{-\infty}^{\infty} d\xi n_F(\xi)[\delta(\xi-\epsilon-\omega_q)-\delta(\xi-\epsilon+\omega_q)]
    \Lambda_{\alpha\beta}^{RA}(\xi,\xi){\cal I}_{\beta\gamma}^{RA}(\xi,\xi; \mathbf k+\mathbf q), 
\end{equation}
where 
\begin{equation}
I^{RA}_{\beta\gamma}(\xi,\xi; \mathbf k+\mathbf q)=\frac{1}{2}\Tr[\sigma_{\beta}G^R(\xi,\mathbf k+\mathbf q)\sigma_{\gamma}G^A(\xi,\mathbf k+\mathbf q)].
\end{equation}

Performing the same analytic continuation for the residue terms in Eq.(\ref{eq:S}) and then taking the limit $\omega\to 0$, we get 
\begin{eqnarray}\label{eq:residue}
&&\sum_{z_j=\pm \omega_q}Res [D_0(z=z_j)]\Lambda_{\alpha\beta}(z_j+\epsilon+\omega+i 0^+, z_j+\epsilon-i 0^+) {\cal I}_{\beta\gamma}(z_j+ \epsilon +\omega + i 0^+, z_j+ \epsilon-i 0^+)n_B(z_j) \nonumber\\
    &&=n_B(\omega_q)\Lambda_{\alpha\beta}^{RA}(\omega_q+\epsilon,\omega_q+\epsilon){\cal I}^{RA}_{\beta\gamma}( \omega_q+\epsilon , \omega_q+\epsilon )
    - n_B(-\omega_q)\Lambda_{\alpha\beta}^{RA}(-\omega_q+\epsilon,-\omega_q+\epsilon){\cal I}^{RA}_{\beta\gamma}( -\omega_q+\epsilon , -\omega_q+\epsilon ) \nonumber\\
    &&= \int_{-\infty}^{\infty} d\xi \ \Lambda_{\alpha\beta}^{RA}(\xi,\xi){\cal I}^{RA}_{\beta\gamma}( \xi, \xi)[\delta(\xi-\epsilon-\omega_q)n_B(\omega_q)
    + \delta(\xi-\epsilon+\omega_q)(1+n_B(\omega_q))].
\end{eqnarray}

From  Eq.(\ref{eq:S}), (\ref{eq:S_real}) and (\ref{eq:residue}), we get 
\begin{equation}
Q^{RA}(\epsilon,\epsilon)
    =\int_{-\infty}^{\infty} d\xi \Lambda_{\alpha\beta}^{RA}(\xi,\xi){\cal I}^{RA}_{\beta\gamma}( \xi, \xi)
    [ \delta(\xi-\epsilon-\omega_q)(n_B(\omega_q)+ n_F(\xi)) +
   \delta(\xi-\epsilon+\omega_q)(n_B(\omega_q)+ 1-n_F(\xi))].
\end{equation}

The recursion Eq.(\ref{eq:vertex_recursion}) after analytic continuation to the real  energy axis becomes 
\begin{eqnarray}
  \Lambda_{\alpha\gamma }^{RA}(\epsilon,\epsilon; \mathbf k)&=& \delta_{\alpha\gamma}+ \sum_{\mathbf q}|g_{\mathbf q}|^2 Q^{RA}(\epsilon,\epsilon) \nonumber\\
  &=& \delta_{\alpha\gamma}+ \sum_{\mathbf q}|g_{\mathbf q}|^2
    \int d\xi \Lambda_{\alpha\beta}^{RA}(\xi,\xi; \mathbf k+ \mathbf q){\cal I}^{RA}_{\beta\gamma}(\xi, \xi; \mathbf k+ \mathbf q) \nonumber\\
    && [ \delta(\xi-\epsilon-\omega_{\mathbf q})(n_B(\omega_{\mathbf q})+ n_F(\xi))
   +\delta(\xi-\epsilon+\omega_{\mathbf q})(n_B(\omega_{\mathbf q})+ 1-n_F(\xi))].
\end{eqnarray}

The recursion equation of the current vertex $\Gamma_\alpha$ after analytic continuation to real energy axis is then
\begin{eqnarray}\label{eq:Vertex_recursion_0}
 \hat{\Gamma}_{\alpha}^{RA}(\epsilon,\epsilon;\mathbf k)
    &=& \hat{j}_{\alpha}+\int d\xi\sum_{\mathbf q} | g_{\mathbf q}|^2 G^A(\xi,\mathbf k+\mathbf q)  \hat{\Gamma}_{\alpha}^{RA}(\xi,\xi; \mathbf k+\mathbf q)G^R(\xi,\mathbf k+\mathbf q) \nonumber\\
    && [ \delta(\xi-\epsilon-\omega_{\mathbf q})(n_B(\omega_{\mathbf q})+ n_F(\xi))
   +\delta(\xi-\epsilon+\omega_{\mathbf q})(n_B(\omega_{\mathbf q})+ 1-n_F(\xi))].
   \end{eqnarray}
   
To lighten the notation, we drop the superscript $RA$ in $ \hat{\Gamma}_{\alpha}^{RA},  \Lambda_{\alpha\gamma }^{RA}$ and 
 ${\cal I}^{RA}_{\beta\gamma}$ and assume we are discussing the $RA$ component of these quantities by default  in the following text. 

\subsection{Renormalized current vertex in the band basis}

The dominant vertex correction comes from the phonon scatterings of electrons within the upper band. It is then convenient to work in the eigenstate band basis (chiral basis) to compute  the dominant vertex correction.

The renormalized current vertex in the Feynman diagrams of the AH conductivities corresponds to the band diagonal matrix element 
\begin{eqnarray}\label{eq:Vertex_recursion}
\Gamma_{\alpha}^{++}(\epsilon,\epsilon;\mathbf k)&\equiv& \dirac {u_{\mathbf k}^+}{ \hat{\Gamma}_{\alpha}(\epsilon,\epsilon;\mathbf k)}  {u_{\mathbf k}^+} \nonumber\\
&=& {j}^{++}_{\alpha}(\mathbf  k) +\int d\xi\sum_{\mathbf  q}  |{g}_{\mathbf  q}|^2 G^{R+}(\xi,\mathbf  k+\mathbf q) G^{A+}(\xi,\mathbf  k+\mathbf q)\Gamma_{\alpha}^{++}(\xi,\xi; \mathbf  k+\mathbf q)  |\inner{u_{\mathbf k+\mathbf q}^+}{u_{\mathbf k}^+}|^2
   \nonumber\\
    &&[ \delta(\xi-\epsilon-\omega_{\mathbf q})(n_B(\omega_{\mathbf q})+ n_F(\xi))
   +\delta(\xi-\epsilon+\omega_{\mathbf q})(n_B(\omega_{\mathbf q})+ 1-n_F(\xi))],
\end{eqnarray}
where ${j}^{++}_{\alpha}(\mathbf k)=\dirac {u_{\mathbf k}^+}{ \hat{j}_{\alpha}}  {u_{\mathbf k}^+}=ev \frac{vk_\alpha}{\epsilon_k}$ and 
\begin{equation}
G^{R/A, +}(\epsilon, \mathbf k)=\dirac{u_{\mathbf k}^+}{ \hat{G}^{R/A}}{u_{\mathbf k}^+} = \frac{1 }{\epsilon -\epsilon_{k}^+ \pm \frac{i}{2\tau_k^+}}.
\end{equation}

The  recursion Eq.(\ref{eq:Vertex_recursion}) of the current vertex is hard to solve exactly. We then apply the approximation that the scattering by phonon is quasi-elastic as before, i.e., $\epsilon_{k'}=\epsilon_k\pm\omega_q\approx \epsilon_k, k'\approx k$ in Eq.(\ref{eq:Vertex_recursion}), where $\mathbf k'\equiv \mathbf k+\mathbf q$. Under this approximation, we can compute the renormalized current vertex $\Gamma_{\alpha}^{++}(\epsilon,\epsilon;\mathbf k)$ order by order by iteration of Eq.(\ref{eq:Vertex_recursion}). In the following, we show this process for 
$\Gamma_{x}^{++}(\epsilon,\epsilon;\mathbf k)$.

The sum over the phonon momentum $\mathbf q$ in Eq.(\ref{eq:Vertex_recursion}) may be replaced by the sum over $\mathbf k'$ as in the calculation of the self-energy and Eq.(\ref{eq:Vertex_recursion}) becomes
\begin{eqnarray}\label{eq:Vertex_recursion_1}
\Gamma_{\alpha}^{++}(\epsilon,\epsilon;\mathbf k)
&=& {j}^{++}_{\alpha}(\mathbf  k) +\int d\xi \int \frac{k'dk'}{(2\pi)^2} \int_0^{2\pi}d\theta  |{g}_{\mathbf  q}|^2 G^{R+}(\xi,\mathbf  k') G^{A+}(\xi,\mathbf  k')\Gamma_{\alpha}^{++}(\xi,\xi; \mathbf  k')  |\inner{u_{\mathbf k'}^+}{u_{\mathbf k}^+}|^2
   \nonumber\\
    &&[ \delta(\xi-\epsilon-\omega_{\mathbf q})(n_B(\omega_{\mathbf q})+ n_F(\xi))
   +\delta(\xi-\epsilon+\omega_{\mathbf q})(n_B(\omega_{\mathbf q})+ 1-n_F(\xi))].
\end{eqnarray}

For $\Gamma^{++}_x$, the zeroth order is ${j}^{++}_{x}(\mathbf  k)=e v^2k_x/\epsilon_k$. The first order can be obtained by  
replacing $\Gamma_{\alpha}^{++}(\xi,\xi; \mathbf  k')$ in Eq.(\ref{eq:Vertex_recursion_1}) with  ${j}^{++}_{x}(\mathbf k')$.
Since $v^2 k' dk'=\epsilon_{k'}d \epsilon_{k'}$, we can replace the integration over $dk'$ by $d\epsilon_{k'}$  in Eq.(\ref{eq:Vertex_recursion_1}). 
Employing
\begin{eqnarray}
G^{R+}(\xi,\mathbf k') G^{A+}(\xi,\mathbf k') &=&2\pi \tau^+_{k'}\delta(\xi-\epsilon_{k'}),\  \tau_{k'}^+={1}/[a + \frac{bv^2k'^2+a\Delta^2}{\xi\epsilon_{k'}}], \\
 |\inner{u_{\mathbf k'}^+}{u_{\mathbf k}^+}|^2 &=&\frac{1}{2}(1+\cos\alpha'\cos\alpha+\sin\alpha'\sin\alpha\cos\theta), \\
 {j}^{++}_{x}(\mathbf k') &=& ev \frac{vk'}{\epsilon_{k'}}\cos(\theta+\theta_0),
\end{eqnarray}
 we get the first order of $\Gamma^{ ++}_x$ after integration over $dk'$ as
\begin{eqnarray}\label{eq:Gamma_x_1}
&&\Gamma_{x}^{(1),++}(\epsilon,\epsilon;\mathbf k)
\approx g^2_D \frac{\hbar}{4\pi\rho s^2 v^2 } \frac{e v^2 k}{\epsilon_k}\tau^+_{k}\int_{-\infty}^{\infty} \xi d\xi  \int_0^{2\pi}d\theta \int_0^{2k_B T_{BG}} \Omega d\Omega \delta(\Omega-\omega_q)   (\cos\theta_0 \cos\theta-\sin\theta_0 \sin\theta)   \nonumber\\
    &&\ \  \frac{1}{2}(1+\cos^2\alpha+\sin^2\alpha\cos\theta)
[ \delta(\xi-\epsilon-\Omega)(n_B(\Omega)+ n_F(\xi))
   +\delta(\xi-\epsilon+\Omega)(n_B(\Omega)+ 1-n_F(\xi))].
\end{eqnarray}
In the above integration, we have applied the quasi-elastic scattering  approximation so that  $\tau^+_{k'}\approx \tau^+_{k}, \cos\alpha'\approx \cos\alpha=\Delta/\epsilon_k, \sin\alpha'\approx \sin\alpha$. 
We also introduced an  integration over $d\Omega$ through the factor $\delta(\Omega-\omega_q)$ to convert the integration over the angle $d\theta$ to the integration over $d\Omega$ as in the calculation of the self-energy. The integration over $d\theta$ can be done using the following integrals
\begin{eqnarray}
 \int_{0}^{2\pi} d \theta \delta(\Omega-\omega_q)&\approx&\frac{2}{sk}\frac{1}{|\cos\frac{\theta_\Omega}{2}|}\approx\frac{2}{sk}\frac{1}{\sqrt{1-\frac{\Omega^2}{4s^2k^2}}} , \ \theta_\Omega\equiv 2 \arcsin\frac{\Omega}{2sk}, \\
    \int_{0}^{2\pi} d \theta \cos\theta \delta(\Omega-\omega_q)&\approx & \frac{2}{sk}\frac{\cos\theta_\Omega}{|\cos\frac{\theta_\Omega}{2}|} \approx \frac{2}{sk}\frac{1-\Omega^2/2s^2k^2}{\sqrt{1-\frac{\Omega^2}{4s^2k^2}}}, \\
     \int_{0}^{2\pi} d \theta \cos^2\theta \delta(\Omega-\omega_q)&\approx& \frac{2}{sk}\frac{\cos^2\theta_\Omega}{|\cos\frac{\theta_\Omega}{2}|} \approx \frac{2}{sk}\frac{(1-\Omega^2/2s^2k^2)^2}{\sqrt{1-\frac{\Omega^2}{4s^2k^2}}}, \\
     \int_{0}^{2\pi} \sin\theta d \theta \delta(\Omega-\omega_q)&=& 0, \ \ \    \int_{0}^{2\pi} \sin\theta \cos\theta d \theta \delta(\Omega-\omega_q)= 0.
\end{eqnarray}

After the integration over $d\theta$ in Eq.(\ref{eq:Gamma_x_1}), we get the first order of $\Gamma^{++}_x$ as
\begin{eqnarray}\label{eq:Gamma_1}
\Gamma_{x}^{(1),++}(\epsilon,\epsilon;\mathbf k)
    &&\approx \frac{ g_D^2 }{4 \pi \rho s^4 v^2 k^2} \frac{e v^2 k_x}{\epsilon_k } \tau^+_k \int_0^{2k_B T_{BG}} d\Omega \Omega^2
   \int_{-\infty}^\infty \xi d\xi \frac{\cos\theta_\Omega}{|\sin \theta_\Omega| }
 \times [(1+\frac{\Delta^2}{\xi\epsilon_k})+ \frac{v^2k^2}{\xi\epsilon_k}\cos\theta_\Omega] \nonumber\\
  && \times [\delta(\xi-\epsilon-\Omega) ((n_B(\Omega)+ n_F(\xi) )
    +\delta(\xi-\epsilon+\Omega)(n_B(\Omega)+ 1-n_F(\xi))] \nonumber\\
  &&\approx \frac{ev^2 k_x}{\epsilon_k} \tau^+_k [ (1+\frac{\Delta^2}{\epsilon\epsilon_k})b(\epsilon,k) + \frac{v^2k^2}{\epsilon\epsilon_k} c(\epsilon,k)],
\end{eqnarray}
where $b(\epsilon, k)$ is defined in Eq.(\ref{eq:b}) and $c(\epsilon, k)$ is defined as 
\begin{eqnarray}\label{eq:c}
c(\epsilon, k)
   & =&\frac{1}{4\pi}\frac{g_D^2}{\rho s^4 v^2}\frac{1}{k^2}
    \int_{-\infty}^\infty \xi d \xi \int_{0}^{k_B T_{BG}}d \Omega \ \Omega^2 \frac{\cos^2 \theta_\Omega}{|\sin \theta_\Omega|}\nonumber\\
   && \times [ \delta(\xi-\epsilon-\Omega)(n_B(\Omega)+n_F(\xi))
      + \delta(\xi-\epsilon+\Omega)(n_B(\Omega)+1-n_F(\xi))] \nonumber\\
 &\approx& \frac{1}{4\pi}\frac{g_D^2 \hbar}{\rho s^3 v^2}\frac{\epsilon}{k}
    \int_{0}^{k_B T_{BG}} \Omega d \Omega \  (1-\frac{\Omega^2}{2s^2 k^2})^2 (1-\frac{\Omega^2}{4s^2 k^2} )^{-\frac{1}{2}}   [2n_B(\Omega)+1+n_F(\epsilon+\Omega)
     -n_F(\epsilon-\Omega))].     
\end{eqnarray}

We denote 
\begin{eqnarray}
\lambda(\epsilon, k)&=&\tau^+_k [ (1+\frac{\Delta^2}{\epsilon\epsilon_k})b(\epsilon,k) + \frac{v^2k^2}{\epsilon\epsilon_k} c(\epsilon,k)]. 
\end{eqnarray}
From Eq.(\ref{eq:Gamma_1}), we get 
\begin{equation}
\Gamma_{x}^{(1),++}(\epsilon,\epsilon;\mathbf k)=\lambda(\epsilon, k) j^{++}_x(\mathbf k).
\end{equation}

By iteration order by order we get 
\begin{equation}
\Gamma_{x}^{(n),++}(\epsilon,\epsilon;\mathbf k)=\lambda^n(\epsilon, k) j^{++}_x(\mathbf k),
\end{equation}
and the renormalized current vertex 
\begin{equation}
\Gamma_{x}^{++}=\sum_{n=0}^{\infty} \Gamma_{x}^{(n),++}(\epsilon,\epsilon;\mathbf k)=\frac{1}{1-\lambda} j^{++}_x(\mathbf k).
\end{equation}

Since the system is isotropic, $\Gamma_{y}^{++}=\sum_{n=0}^{\infty} \Gamma_{y}^{(n),++}(\epsilon,\epsilon;\mathbf k)=\frac{1}{1-\lambda} j^{++}_y(\mathbf k)$.

We have checked that the current vertex renormalization factor $\gamma\equiv \frac{1}{1-\lambda}$ is equal to $\tau_k^{tr}/\tau_k^+$ where $\tau_k^{tr}$ and $\tau_k^+$ are respectively the transport and mean lifetime of the upper band electrons with phonon scatterings defined in Ref.~\cite{Niu2019} as
\begin{eqnarray}
1/\tau^+_k&=&\sum_{\mathbf k'}\omega^{(2)}_{\mathbf k, \mathbf k'}\frac{1-f^0_{\mathbf k'}}{1-f^0_{\mathbf k}}, \label{eq:SC_tau_+}\\
1/\tau_k^{tr}&=&\sum_{\mathbf k'}\omega^{(2)}_{\mathbf k, \mathbf k'}\frac{1-f^0_{\mathbf k'}}{1-f^0_{\mathbf k}}(1-\cos \phi_{\mathbf k', \mathbf k}),
\end{eqnarray}
where $\omega^{(2)}_{\mathbf k, \mathbf k'}=2\pi |g_{\mathbf k' \mathbf k}|^2 |\inner{u_{\mathbf k'}^+}{u_{\mathbf k}^+}|^2 
    [ \delta(\epsilon_{k'}-\epsilon_k-\omega_q)n_B(\omega_q) + \delta(\epsilon_{k'}-\epsilon_k+\omega_q)(n_B(\omega_q)+1)]$  is the 2nd order e-phonon scattering rate from $\mathbf k$ to $\mathbf k'$, $\phi_{\mathbf k', \mathbf k}$ is the angle between $\mathbf k$ and $\mathbf k'$ and $f^0_{\mathbf k}$ is the Fermi distribution function for energy $\epsilon_k$. Note that  $1/\tau^+_k$ defined in Eq.(\ref{eq:SC_tau_+}) is also equal to that in Eq.(\ref{eq:tau^pm}).

\section{Appendix C: Anomalous Hall conductivity}
The extrinsic contribution of the dc AH conductivity comes from $\sigma_{xy}^{\rm I}$ which can be written as 
\begin{equation}\label{eq:sigma_I}
  \sigma^{\rm I}_{xy}= e^2v^2\sum_{\mathbf k}\int\frac{d\epsilon}{2\pi}(-\partial_{\epsilon} n_F(\epsilon)) \Tr[\hat{\Gamma}_{x}(\epsilon,\epsilon,\mathbf k)G^R(\epsilon,\mathbf k)\sigma_{y}G^A(\epsilon,\mathbf k)].
\end{equation}
Since $\partial_{\epsilon} n_F(\epsilon)\sim \delta{(\epsilon-\epsilon_F)}$, the contribution to $\sigma^{\rm I}_{xy}$ comes from the electrons on the Fermi surface.

The  contribution to $\sigma^{\rm I}_{xy}$ can be separated to three parts due to different mechanisms: the intrinsic, the side jump and the skew scattering contributions. The intrinsic contribution is due to the non-trivial band structure of the clean system and has been calculated in previous works for 2D massive Dirac metals~\cite{Sinitsyn2007}. The side jump and skew scattering contributions can be most easily separated by expanding the trace in Eq.(\ref{eq:sigma_I}) in the chiral basis, as shown in our previous work~\cite{Zhang2023}. The resulting AH conductivities are depicted by the Feynman diagrams in the chiral basis in Fig.\ref{fig:AHE_diagram} of the main text or Fig.\ref{fig:side_jump} and Fig.\ref{fig:skew_scattering} in the appendix.

\subsection{Side jump contribution}

We first calculate the side jump contribution. The dc AH conductivity from Fig.\ref{fig:side_jump}(a) and (b)  can be written as
\begin{equation}\label{eq:side_jump_1}
 \sigma^{a+b}_{xy}= \sum_{\mathbf k}\int\frac{d\epsilon}{2\pi}(-\partial_{\epsilon} n_F(\epsilon)) [\Gamma'^{+-}_{x}(\epsilon, \epsilon; \mathbf k) G_0^{R-}(\epsilon, \mathbf k) j_{y}^{-+}(\mathbf k)G_0^{A+}(\epsilon, \mathbf k) + \Gamma'^{-+}_{x}(\epsilon, \epsilon; \mathbf k) G_0^{R+}(\epsilon, \mathbf k) j_{y}^{+-}(\mathbf k)G_0^{A-}(\epsilon, \mathbf k)],
\end{equation}
where
$\hat{\Gamma}'_x\equiv \hat{\Gamma}_x-\hat{j}_x$ and $\Gamma'^{+-}_{x}\equiv \langle u_{\mathbf k}^+| \hat{\Gamma}'_x | u_{\mathbf k}^-\rangle$.

Applying the recursion Eq.(\ref{eq:Vertex_recursion_0}) of the renormalized current vertex $\hat{\Gamma}_x$, we get
\begin{eqnarray}\label{eq:Gamma'}
  \Gamma'^{+-}_{x}(\epsilon,\epsilon;\mathbf k)
   & =& \int d\xi\sum_{\mathbf k'}  |{g}_{\mathbf k'-\mathbf k}|^2 G^{R+}(\xi,\mathbf k') G^{A+}(\xi,\mathbf k')\Gamma_x^{++}(\xi,\xi; \mathbf k')  \inner{u_{\mathbf k}^+}{u_{\mathbf k'}^+} \inner{u_{\mathbf k'}^+}{u_{\mathbf k}^-}
   \nonumber\\
    &&\times    [ \delta(\xi-\epsilon-\omega_{\mathbf k'-\mathbf k})(n_B(\omega_{\mathbf k'-\mathbf k})+ n_F(\xi))
    +\delta(\xi-\epsilon+\omega_{\mathbf k-\mathbf k'})(n_B(\omega_{\mathbf k'-\mathbf k})+ 1-n_F(\xi))],
\end{eqnarray}
 where 
 \begin{eqnarray}
  &&\inner{u_{\mathbf k}^+}{u_{\mathbf k'}^+}=\cos\frac{\alpha'}{2}\cos\frac{\alpha}{2}+\sin\frac{\alpha'}{2}\sin\frac{\alpha}{2}e^{i\theta},
\ \  \ \inner{u_{\mathbf k'}^+}{u_{\mathbf k}^-}=\cos\frac{\alpha'}{2}\sin\frac{\alpha}{2}-\sin\frac{\alpha'}{2}\cos\frac{\alpha}{2}e^{-i\theta},\\
&& \ \   G^{R+}(\xi,\mathbf k') G^{A+}(\xi,\mathbf k')=2\pi \tau^+_{k'}\delta(\xi-\epsilon_{k'}), \ \ \ {\rm and} \ \ \ \Gamma_{x}^{++}(\epsilon,\epsilon; \mathbf k)= \frac{1}{1-\lambda}\frac{e v^2 k_x}{\epsilon}.
 \end{eqnarray}
The sum over $\mathbf k'$ in Eq.(\ref{eq:Gamma'}) can be done by the same procedure as for the calculation of $\Gamma_x^{(1), ++}$ under the quasi-elastic scattering approximation. We get
\begin{equation}\label{eq:Gamma'_x_pm}
 \Gamma'^{+-}_{x}(\epsilon,\epsilon;\mathbf k)=\frac{\tau^+_k}{1-\lambda(\epsilon)} \frac{v^2k}{\epsilon^2}\left[\frac{\Delta}{\epsilon_k} (b(\epsilon,k)- c(\epsilon,k)) k_x-i d(\epsilon,k)k_y \right],
\end{equation}
 where $d(\epsilon, k)=a(\epsilon, k)-c(\epsilon, k)$. Similarly, we get 
 \begin{equation}\label{eq:Gamma'_y_pm}
 \Gamma'^{+-}_{y}(\epsilon,\epsilon;\mathbf k)=\frac{\tau^+_k}{1-\lambda(\epsilon)} \frac{v^2k}{\epsilon^2}\left[\frac{\Delta}{\epsilon_k} (b(\epsilon,k)- c(\epsilon,k)) k_y+ i d(\epsilon,k)k_x \right].
\end{equation}

\begin{figure}
	\includegraphics[width=15cm]{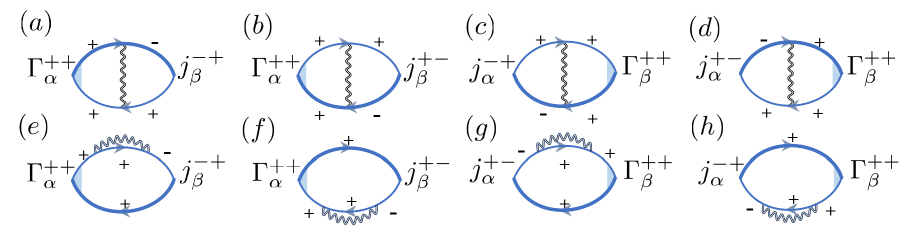}
			\caption{(a)Feynman diagrams of the side jump conductivity in the chiral basis. The thin and thick solid lines represent  the bare electron GF and the electron GF in the first Born approximation respectively. The curvy  lines represent the phonon propagators. Note that replacing the thin solid lines by the thick ones in the diagrams, as shown in Fig.\ref{fig:AHE_diagram} of the main text,  does not change the AH conductivity of the diagrams.}\label{fig:side_jump}
\end{figure}

With the above ingredients, we can compute the AH conductivity in Eq.(\ref{eq:side_jump_1}) corresponding to the Feynman diagrams in Fig.\ref{fig:side_jump}(a) and (b):
\begin{equation}\label{eq:side_jump_1'}
\sigma^{a+b}_{xy}= \frac{1}{2\pi}\sum_{\mathbf k}  [\Gamma'^{+-}_{x}(\epsilon, \epsilon; \mathbf k) G_0^{R-}(\epsilon, \mathbf k) j_{y}^{-+}(\mathbf k)G_0^{A+}(\epsilon, \mathbf k) + \Gamma'^{-+}_{x}(\epsilon, \epsilon; \mathbf k) G_0^{R+}(\epsilon, \mathbf k) j_{y}^{+-}(\mathbf k)G_0^{A-}(\epsilon, \mathbf k)] |_{\epsilon=\epsilon_F},
\end{equation}
where 
\begin{eqnarray}
&& j_y^{-+}(\mathbf k)=ev  \sigma_{y}^{-+}(\mathbf k), \ \ \ \ \ \   \sigma_{y}^{-+}(\mathbf k)= -i\cos\theta_0 -\cos\alpha \sin\theta_0, \\
 &&   G_0^{R-}(\epsilon,k)=\frac{1}{\epsilon-\epsilon_k^- + i\eta}, \ \ \ \ \ \
    G_0^{A+}(\epsilon,k)=\frac{1}{\epsilon-\epsilon_k^+ -i\eta}, \ \  \eta \rightarrow 0^+, \\
 &&  \Gamma'^{+-}_{x} j_{y}^{-+} =[\Gamma'^{-+}_{x} j_{y}^{+-}]^* = e^2 v^2\frac{i \tau^+_k }{1-\lambda(\epsilon)} \frac{v^2k^2}{\epsilon^2}[ - (b(\epsilon,k)- c(\epsilon,k)) \cos\alpha\cos^2\theta_0 + d(\epsilon,k)\cos\alpha\sin^2\theta_0].
\end{eqnarray}

The sum over $\mathbf k$ in Eq.(\ref{eq:side_jump_1'}) may be converted to the integral
\begin{equation}
\sum_{\mathbf k}\to \int_0^{+\infty} \frac{k dk}{(2\pi)^2} \int_0^{2\pi} d\theta_0.
\end{equation}
After the integration over $d \theta_0$, we get
\begin{eqnarray}
\sigma^{a+b}_{xy}&=&\frac{i}{2} \frac{\Delta}{\epsilon^2} \int \frac{k dk}{4\pi^2} \frac{\tau^+_k }{1-\lambda(\epsilon)} \frac{e^2 v^4k^2}{\epsilon_k}[ a(\epsilon,k) -b(\epsilon,k)]
     ( \frac{1}{\epsilon-\epsilon_k^- + i\eta} \frac{1}{\epsilon-\epsilon_k^+ -i\eta}-\frac{1}{\epsilon-\epsilon_k^+ + i\eta} \frac{1}{\epsilon-\epsilon_k^- -i\eta}  )|_{\epsilon=\epsilon_F} \nonumber\\
     &=&\frac{i}{8\pi^2} \frac{\Delta}{\epsilon^2} \int k dk \frac{\tau^+_k}{1-\lambda(\epsilon)} \frac{e^2 v^4 k^2}{\epsilon_k}[ a(\epsilon,k) -b(\epsilon,k)]\frac{1}{\epsilon+\epsilon_k^+}\times 2i\pi \delta(\epsilon-\epsilon_k^+)|_{\epsilon=\epsilon_F} \nonumber\\
     &=&-\frac{e^2}{8\pi }\frac{\Delta}{\epsilon_F} (1-\frac{\Delta^2}{\epsilon_F^2})\frac{\tau^+_{k_F}}{1-\lambda(\epsilon_F,k_F)}[a(\epsilon_F,k_F)-b(\epsilon_F,k_F)].
\end{eqnarray}

The total contribution from the diagrams  Fig.\ref{fig:side_jump}(c) and (d) is identical to that of Fig.\ref{fig:side_jump}(a) and (b). We then get the total side jump contribution due to Fig.\ref{fig:side_jump} (a)-(d) as 
\begin{equation}
\sigma^{side, (1)}_{xy}=-\frac{e^2}{4\pi }\frac{\Delta}{\epsilon_F} (1-\frac{\Delta^2}{\epsilon_F^2})\frac{\tau^+_{k_F}}{1-\lambda(\epsilon_F,k_F)}[a(\epsilon_F,k_F)-b(\epsilon_F,k_F)].
\end{equation}

We next compute the contribution from Fig.\ref{fig:side_jump}(e). The AH conductivity corresponding to this diagram can be written as
\begin{equation}\label{eq:side_jump_e}
\sigma^e_{xy}=\frac{1}{2\pi}\sum_{\mathbf k} \Gamma_x^{++}G_0^{R+}\Sigma^{R,+-}G_0^{R-}j_y^{-+}G^{A+}|_{\epsilon=\epsilon_F},
\end{equation}
where
\begin{equation}
\Sigma^{R,+-}=\dirac{u^+_{\mathbf k}}{\Sigma^{R}}{u^-_{\mathbf k}}=-\frac{i}{2}\frac{vk\Delta}{\epsilon\epsilon_k}(a-b).
\end{equation}
 After the sum over $\mathbf k$ in Eq.(\ref{eq:side_jump_e}), we get 
\begin{equation}
\sigma^e_{xy}=-\frac{e^2}{16\pi}\frac{\Delta}{\epsilon_F}(1-\frac{\Delta^2}{\epsilon_F^2})\frac{\tau^+_{k_F}}{1-\lambda(\epsilon_F, k_F)}(a(\epsilon_F, k_F)-b(\epsilon_F, k_F)).
\end{equation}

Each of the diagrams (f)-(h) contributes the same as diagram (e) in Fig.\ref{fig:side_jump} so the total contribution from the diagrams (e)-(h) is
\begin{equation}
\sigma^{side, (2)}_{xy}=-\frac{e^2}{4\pi }\frac{\Delta}{\epsilon_F}(1-\frac{\Delta^2}{\epsilon_F^2})\frac{\tau^+_{k_F}}{1-\lambda(\epsilon_F, k_F)}(a(\epsilon_F, k_F)-b(\epsilon_F, k_F)).
\end{equation}

The total side jump conductivity is 
\begin{eqnarray}\label{eq:side_jump}
\sigma^{side}_{xy}&=&\sigma^{side, (1)}_{xy}+\sigma^{side, (2)}_{xy} \nonumber\\
&=& -\frac{e^2}{2\pi }\frac{\Delta}{\epsilon_F}(1-\frac{\Delta^2}{\epsilon_F^2})\frac{\tau^+_{k_F}}{1-\lambda(\epsilon_F, k_F)}(a(\epsilon_F, k_F)-b(\epsilon_F, k_F)) \nonumber\\
&=&-\frac{e^2}{2\pi }\frac{\Delta}{\epsilon_F}(1-\frac{\Delta^2}{\epsilon_F^2})\frac{a(\epsilon_F, k_F)-b(\epsilon_F, k_F)}{a(\epsilon_F, k_F)-c(\epsilon_F, k_F)+\frac{\Delta^2}{\epsilon_F^2}\left[a(\epsilon_F, k_F)+c(\epsilon_F, k_F)-2b(\epsilon_F, k_F)\right]}.
\end{eqnarray}

\subsection{Intrinsic Skew scattering contribution}

The intrinsic skew scattering contribution from non-crossing diagrams is described by the Feynman diagrams in Fig.\ref{fig:skew_scattering}. The AH conductivity due to diagrams (a) and (b) in Fig.\ref{fig:skew_scattering} can be written as 
\begin{eqnarray}
\sigma_{xy}^{sk, a+b}=\sum_{\mathbf k}\int\frac{d\epsilon}{2\pi}(-\partial_{\epsilon} n_F(\epsilon))[\Gamma'^{+-}_x G^{R-}\Gamma'^{-+}_y G_0^{A+} + \Gamma'^{-+}_x G_0^{R+}\Gamma'^{+-}_y G^{A-}],
\end{eqnarray}
where $\Gamma'^{+-}_x=[\Gamma'^{-+}_x]^*, \Gamma'^{+-}_y=[\Gamma'^{-+}_y]^*$ and $\Gamma'^{+-}_x, \Gamma'^{+-}_y$ are given in Eq.(\ref{eq:Gamma'_x_pm}) and (\ref{eq:Gamma'_y_pm}). And $G_0$ and $G$ are the bare electron GF and the GF in the first Born approximation respectively, both of which are given in the previous text. After the sum over $\mathbf k$, we get  
\begin{equation}
\sigma_{xy}^{sk, a+b}= \frac{e^2\Delta}{4\pi\epsilon_F}(1-\frac{\Delta^2}{\epsilon_F^2})^2 \left[\frac{\tau^+_{k_F}}{1-\lambda(\epsilon_F,k_F)}\right] ^2\left[b(\epsilon_F,k_F)- c(\epsilon_F,k_F)\right]d(\epsilon_F,k_F).
\end{equation}

\begin{figure}
	\includegraphics[width=15cm]{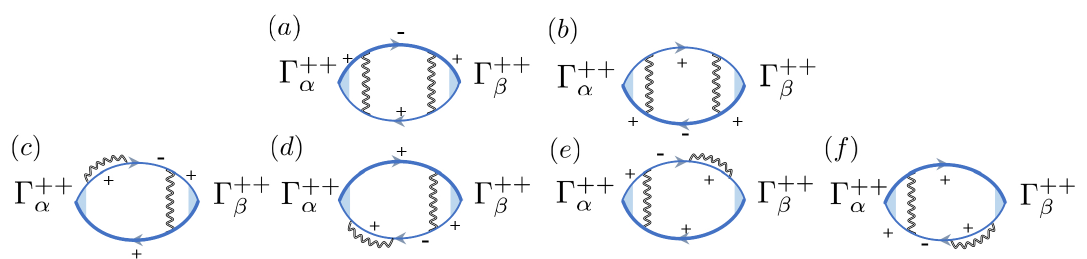}
			\caption{Feynman diagrams of the intrinsic skew scattering conductivity in the chiral basis. The notations are the same as in Fig.\ref{fig:side_jump}.}\label{fig:skew_scattering}
\end{figure}

The AH conductivity due to diagram (c) in Fig.\ref{fig:skew_scattering} can be written as 
\begin{eqnarray}
\sigma_{xy}^{sk, c}&=&\int\frac{d\epsilon}{2\pi}(-\partial_{\epsilon} n_F(\epsilon)) \sum_{\mathbf k} \Gamma_x^{++} G_0^{R+}\Sigma^{R,+-}G_0^{R-}\Gamma^{'-+}_y G^{A+} \nonumber\\
    &&=
    -\frac{e^2\Delta}{16\pi\epsilon_F}(1-\frac{\Delta^2}{\epsilon_F^2})^2 \left[\frac{\tau^+_{k_F}}{1-\lambda(\epsilon_F,k_F)}\right]^2\left[a(\epsilon_F,k_F)- b(\epsilon_F,k_F)\right]d(\epsilon_F,k_F)
 \end{eqnarray}
    
The contribution from each diagram of Fig.\ref{fig:skew_scattering}(e)-(f) is identical to that of (c). The total skew scattering contribution of Fig.\ref{fig:skew_scattering}(a)-(f) is then
\begin{eqnarray}\label{eq:skew_scattering}
\sigma_{xy}^{sk-nc}
    &=&  -\frac{e^2\Delta}{4\pi\epsilon_F}\left(1-\frac{\Delta^2}{\epsilon_F^2}\right)^2 \left[\frac{\tau^+_{k_F}}{1-\lambda(\epsilon_F,k_F)}\right] ^2 [a(\epsilon_F,k_F)+c(\epsilon_F,k_F)- 2b(\epsilon_F,k_F)]d(\epsilon_F,k_F) 
\end{eqnarray}
where $\frac{\tau^+_{k_F}}{1-\lambda(\epsilon_F,k_F)}=\{a(\epsilon_F, k_F)-c(\epsilon_F, k_F)+\frac{\Delta^2}{\epsilon_F^2}\left[a(\epsilon_F, k_F)+c(\epsilon_F, k_F)-2b(\epsilon_F, k_F)\right]\}^{-1}$ and $d(\epsilon_F,k_F)=a(\epsilon_F,k_F)-c(\epsilon_F,k_F)$.

\subsection{Coherent skew scattering contribution}
In this appendix we show that the phonon-induced coherent skew scattering contribution is exactly opposite to the intrinsic skew scattering contribution in the leading order expansion of the phonon momentum at low temperature, and both contributions vanish as $\sim T^2$ as the temperature approaches zero.

 \begin{figure}[b]
	\includegraphics[width=16cm]{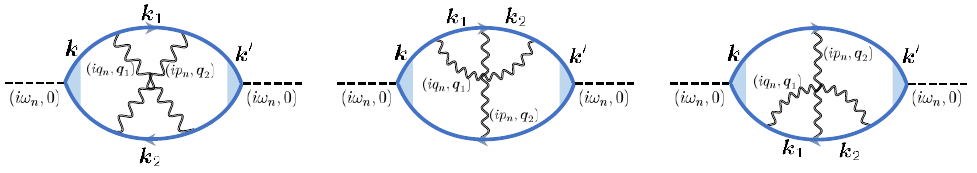}
			\caption{Feynman diagrams of the coherent skew scattering conductivity in the spin basis. The notations are the same as in Fig.\ref{fig:side_jump}.}\label{fig:crossed_spin_basis}
\end{figure}

 The crossed $X$ and $\Psi$ diagrams in the spin basis are shown in Fig.\ref{fig:crossed_spin_basis}. We first calculate the response function of the $X$ diagram, which can be written as 
 \begin{eqnarray}\label{eq:X_response}
 \Pi_{\alpha\beta}^{X}(i\omega_n,\mathbf q=0)
    &=& -\frac{1}{\beta^3}\sum_{\mathbf k}\sum_{\mathbf q_1}\sum_{\mathbf q_2} \sum_{i q_n} \sum_{i p_n} \sum_{i k_n} D_0( i q_n, \mathbf q_1) D_0(i p_n, \mathbf q_2) \nonumber\\
    && {\rm Tr} [G(i k_n, \mathbf k) \Gamma^{\alpha}(i k_n, i k_n+i\omega_n) G(i k_n+i\omega_n, \mathbf k) g_{\mathbf q_1}  G( i k_n+i\omega_n+ iq_n, \mathbf k+\mathbf q_1) g_{\mathbf q_2 } \nonumber\\
    && G( ik_n+i\omega_n+ iq_n+ ip_n, \mathbf k+\mathbf q_1+\mathbf q_2) 
    \Gamma^{\beta}(i k_n+i\omega_n+ i q_n+ i p_n , i k_n+ i q_n+ i p_n)  \nonumber\\
 && G( i k_n+ i q_n+ i p_n, \mathbf k+\mathbf q_1+\mathbf q_2) g^{\dagger}_{\mathbf q_1}  G( i k_n+i p_n, \mathbf k+\mathbf q_2) g^{\dagger}_{\mathbf q_2}].
\end{eqnarray}

For brevity, we denote
\begin{equation}
    \hat{\Upsilon}^{\alpha}(ik_n, ik_n+i\omega_n; \mathbf k)\equiv G(\mathbf k, ik_n) \hat{\Gamma}^{\alpha}(ik_n, ik_n+i\omega_n) G(\mathbf k, ik_n+i\omega_n).
\end{equation}
The response function Eq.(\ref{eq:X_response}) can then be written as 
\begin{eqnarray}\label{eq:X_spin_basis2}
 \Pi_{\alpha\beta}^{X}(i\omega_n, \mathbf q=0)
   & =&-\frac{1}{\beta^3}\sum_{\mathbf k}\sum_{\mathbf q_1}\sum_{\mathbf q_2} \sum_{i q_n} \sum_{i p_n} \sum_{i k_n} |g_{\mathbf q_1}|^2 |g_{\mathbf q_2}|^2
    D(iq_n, \mathbf q_1) D(ip_n, \mathbf q_2){\rm Tr}[ \hat{\Upsilon}^{\alpha}(ik_n, ik_n+i\omega_n; \mathbf k) \nonumber\\
    &&
   G( ik_n+i\omega_n+ iq_n; \mathbf k_1)  \hat{\Upsilon}^{\beta}(ik_n+i\omega_n+ iq_n+ ip_n , ik_n+ iq_n+ ip_n;\mathbf k') G(ik_n+ip_n, \mathbf k_2 ) ],
\end{eqnarray}
where $\mathbf k_1=\mathbf k+\mathbf q_1, \mathbf k_2=\mathbf k+\mathbf q_2, \mathbf k'=\mathbf k+\mathbf q_1 +\mathbf q_2$.

We next expand the trace in the above equation in the band or the chiral basis, and keep only the leading order terms. We get  
\begin{eqnarray}\label{eq:X_chiral_basis}
    \Pi_{\alpha\beta}^{X}(i\omega_n, \mathbf q=0)
    &=&-\frac{1}{\beta^3}\sum_{\mathbf k}\sum_{\mathbf q_1}\sum_{\mathbf q_2} \sum_{iq_n} \sum_{ip_n} \sum_{ik_n} |g_{\mathbf q_1}|^2 |g_{\mathbf q_2}|^2
    D(iq_n, \mathbf q_1) D(ip_n, \mathbf q_2) \nonumber\\
    &&  \Upsilon_{\alpha}^{++}(ik_n, ik_n+i\omega_n; \mathbf k) \Upsilon_{\beta}^{++}(ik_n+i\omega_n+ iq_n+ ip_n , ik_n+ iq_n+ ip_n; \mathbf k') \nonumber\\
    &&[\inner{u_{\mathbf k}^+}{u_{\mathbf k_1}^+}\inner{u_{\mathbf k_1}^+}{u_{\mathbf k'}^+}\inner{u_{\mathbf k'}^+}{u_{\mathbf k_2}^-} \inner{u_{\mathbf k_2}^-}{u_{\mathbf k}^+}  G^{+}( ik_n+i\omega_n+ iq_n, \mathbf k_1)  G^{-}(ik_n+ip_n; \mathbf k_2) \nonumber\\
    &&+\inner{u_{\mathbf k}^+}{u_{\mathbf k_1}^-}\inner{u_{\mathbf k_1}^-}{u_{\mathbf k'}^+}\inner{u_{\mathbf k'}^+}{u_{\mathbf k_2}^+} \inner{u_{\mathbf k_2}^+}{u_{\mathbf k}^+}   G^{-}( ik_n+i\omega_n+ iq_n; \mathbf k_1) G^{+}(ik_n+ip_n; \mathbf k_2)],
\end{eqnarray}
where $\Upsilon_{\alpha}^{++}(ik_n, ik_n+i\omega_n; \mathbf k)\equiv G^+ (ik_n, \mathbf k) \Gamma_{\alpha}^{++}(ik_n, ik_n+i\omega_n) G^+(ik_n+i\omega_n, \mathbf k)$ and $\mathbf k_1, \mathbf k_2, \mathbf k'$ are the same as in Eq.(\ref{eq:X_spin_basis2}). Equation (\ref{eq:X_chiral_basis}) corresponds to the Feynman diagrams in Fig.\ref{fig:crossed_diagrams}(a) and (b).

 \begin{figure}
	\includegraphics[width=16cm]{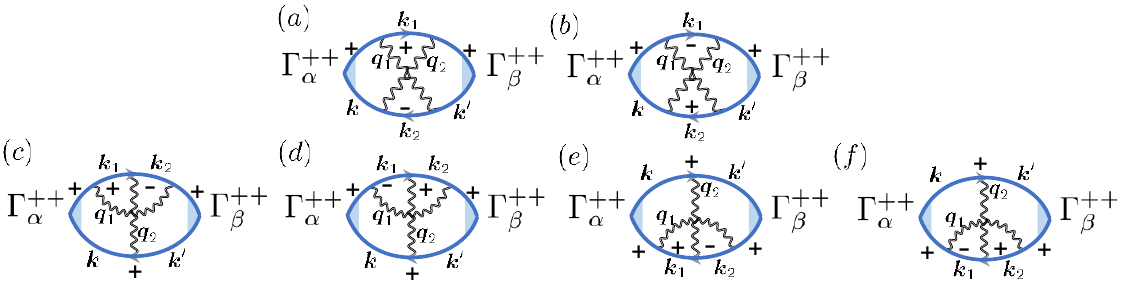}
			\caption{Feynman diagrams of the coherent skew scattering conductivity in the chiral basis. The notations are the same as in Fig.\ref{fig:side_jump}.}\label{fig:crossed_diagrams}
\end{figure}

 Denoting 
 \begin{equation}
 M_X(\mathbf k,\mathbf q_1,\mathbf q_2)\equiv \inner{u_{\mathbf k}^+}{u_{\mathbf k_1}^+}\inner{u_{\mathbf k_1}^+}{u_{\mathbf k'}^+}\inner{u_{\mathbf k'}^+}{u_{\mathbf k_2}^-} \inner{u_{\mathbf k_2}^-}{u_{\mathbf k}^+} ,
 \ \  N_X(\mathbf k,\mathbf q_1,\mathbf q_2)\equiv \inner{u_{\mathbf k}^+}{u_{\mathbf k_1}^-}\inner{u_{\mathbf k_1}^-}{u_{\mathbf k'}^+}\inner{u_{\mathbf k'}^+}{u_{\mathbf k_2}^+} \inner{u_{\mathbf k_2}^+}{u_{\mathbf k}^+},
\end{equation} 
     Eq.(\ref{eq:X_chiral_basis}) can be written as
\begin{eqnarray}\label{eq:X_chiral_basis2}
 \Pi_{\alpha\beta}^{X}(i\omega_n, \mathbf q)
   &=& -\frac{1}{\beta^3}\sum_{\mathbf k}\sum_{\mathbf q_1}\sum_{\mathbf q_2}  |g_{\mathbf q_1}|^2 |g_{\mathbf q_2}|^2 M_X(\mathbf k,\mathbf q_1,\mathbf q_2) \nonumber\\
   && \sum_{ik_n}\sum_{iq_n} \sum_{iQ_n} D(iq_n, \mathbf q_1) D(iQ_n-iq_n, \mathbf q_2) \Upsilon^{++}_{\alpha}(ik_n, ik_n+i\omega_n; \mathbf k)\Upsilon^{++}_{\beta}(ik_n+i\omega_n+ iQ_n , ik_n+ iQ_n; \mathbf k') \nonumber\\
  && G^{+}( ik_n+i\omega_n+ iq_n; \mathbf k_1)  G^{-}(ik_n+iQ_n-iq_n; \mathbf k_2) +(M_X\leftrightarrow N_X,+ \leftrightarrow -), 
\end{eqnarray} 
where $iQ_n\equiv iq_n+i p_n$, and $+ \leftrightarrow -$ is applied only to $G^+$ and $G^-$ but not on $\Upsilon_\alpha^{++}$.

\begin{figure}
	\includegraphics[width=16cm]{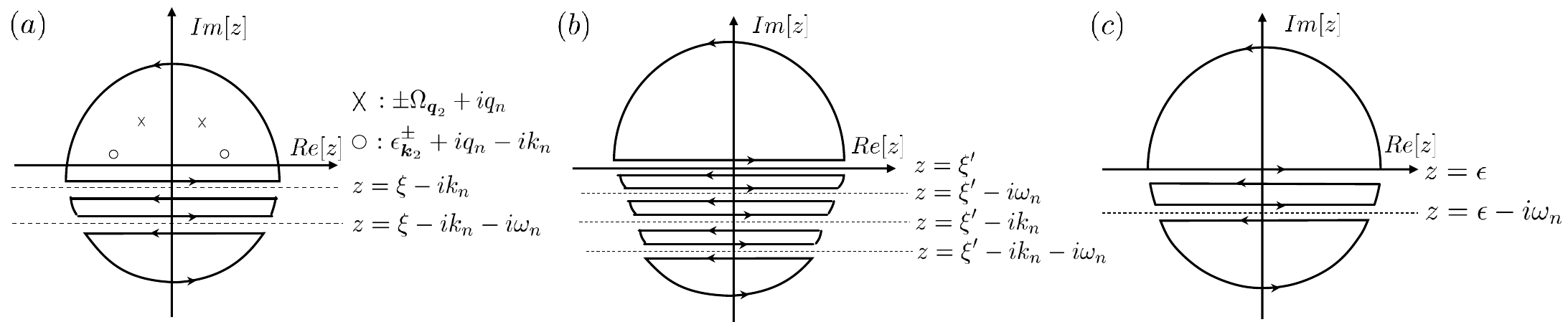}
			\caption{Integration contour for the sum of the Matsubara frequency $iQ_n, i q_n, i k_n$ in Eq.(\ref{eq:X_chiral_basis2}) respectively.}\label{fig:X_contour}
\end{figure}

The sum over the Matsubara frequency $iQ_n, iq_n$ and $ik_n$ in Eq.(\ref{eq:X_chiral_basis2}) can be performed one by one with the method in the previous sections employing the integration contour in Fig.\ref{fig:X_contour}(a), (b) and (c) respectively. After the sum over the Matsubara frequencies and keeping only the leading order terms, we get 
\begin{eqnarray}
 &&\Pi_{\alpha\beta}^{X}(i\omega_n\rightarrow \omega +i0^+)=\frac{i\omega}{2\pi} \sum_{\mathbf k}\sum_{\mathbf q_1}\sum_{\mathbf q_2}  |g_{\mathbf q_1}|^2 |g_{\mathbf q_2}|^2 M_X(\mathbf k,\mathbf q_1,\mathbf q_2) \int d\epsilon [-\partial_{\epsilon} n_F(\epsilon)]\Upsilon_{\alpha}^{AR}(\epsilon, \mathbf k)  \nonumber\\
    &&
    \{ \Upsilon_{\beta}^{RA}(\epsilon+\omega_{\mathbf q_1}+\omega_{\mathbf q_2},  \mathbf k') G^{R+}(\epsilon+\omega_{\mathbf q_1}, \mathbf k_1)G^{A-}(\epsilon+\omega_{\mathbf q_2}, \mathbf k_2)
   [n_B(\omega_{\mathbf q_1})+n_F(\epsilon+\omega_{\mathbf q_1})] [ n_B(\omega_{\mathbf q_2})+n_F(\epsilon+\omega_{\mathbf q_1}+\omega_{\mathbf q_2}) ] \nonumber\\
    &&+  \Upsilon_{\beta}^{RA}(\epsilon+\omega_{\mathbf  q_1}-\omega_{\mathbf  q_2}, \mathbf k') G^{R+}(\epsilon+\omega_{\mathbf  q_1}, \mathbf k_1)G^{A-}(\epsilon-\omega_{\mathbf  q_2}, \mathbf k_2)[n_B(\omega_{\mathbf  q_1})+n_F(\epsilon+\omega_{\mathbf  q_1})] [1+n_B(\omega_{\mathbf  q_2})-n_F(\epsilon+\omega_{\mathbf  q_1}-\omega_{\mathbf  q_2})] \nonumber\\
    &&+ \Upsilon_{\beta}^{RA}(\epsilon-\omega_{\mathbf q_1}+\omega_{\mathbf  q_2}, \mathbf k')G^{R+}(\epsilon-\omega_{\mathbf  q_1}, \mathbf k_1)G^{A-}(\epsilon+\omega_{\mathbf  q_2}, \mathbf k_2)[1+n_B(\omega_{\mathbf  q_1})-n_F(\epsilon-\omega_{\mathbf  q_1})][n_B(\omega_{\mathbf  q_2})+n_F(\epsilon-\omega_{\mathbf q_1}+\omega_{\mathbf  q_2})] \nonumber\\
    &&\ \ \ \ + \Upsilon_{\beta}^{RA}(\epsilon-\omega_{\mathbf q_1}-\omega_{\mathbf  q_2}, \mathbf k')G^{R+}(\epsilon-\omega_{\mathbf  q_1}, \mathbf k_1)G^{A-}(\epsilon-\omega_{\mathbf  q_2}, \mathbf k_2) \nonumber\\
    &&\ \ \ \ \ \  \ [1+n_B(\omega_{\mathbf  q_1})- n_F(\epsilon-\omega_{\mathbf  q_1})][1+n_B(\omega_{\mathbf  q_2})-n_F(\epsilon-\omega_{\mathbf  q_1}-\omega_{\mathbf  q_2}) ] \}  +( M_X\leftrightarrow N_X, +\leftrightarrow -),
\end{eqnarray}
where $\Upsilon_{\alpha}^{AR}(\epsilon, \mathbf k)\equiv G^{A+}(\epsilon, \mathbf k)\Gamma^{++}(\epsilon, \epsilon)G^{R+}(\epsilon, \mathbf k)$.

The above equation can also be written as
\begin{eqnarray}\label{eq:X_Response}
&&\Pi_{\alpha\beta}^{X}(\omega\to 0)
     =\frac{i\omega}{2\pi} \sum_{\mathbf k}\sum_{\mathbf q_1}\sum_{\mathbf q_2}  |g_{\mathbf q_1}|^2 |g_{\mathbf q_2}|^2 M_X(\mathbf k, \mathbf q_1,\mathbf q_2) \int d\epsilon [-\partial_{\epsilon} n_F(\epsilon)]\Upsilon_{\alpha}^{AR}(\epsilon, \mathbf k) \nonumber\\
     &&
      \int d\epsilon_1 [ \delta(\epsilon_1 -\epsilon-\omega_{\mathbf q_1})(n_B(\omega_{\mathbf q_1})+n_F(\epsilon_1)) + \delta(\epsilon_1 -\epsilon+\omega_{\mathbf q_1})(1+n_B(\omega_{\mathbf q_1})-n_F(\epsilon_1))] G^{R+}(\epsilon_1, \mathbf k_1) \nonumber\\
     &&\ \ \ \ \ \ \ \ [ \Upsilon_{\beta}^{RA}(\epsilon_1+\omega_{\mathbf q_2}, \mathbf k')G^{A-}(\epsilon+\omega_{\mathbf q_2}, \mathbf k_2)( n_B (\omega_{\mathbf q_2})+ n_F(\epsilon_1+\omega_{\mathbf q_2})) \nonumber\\
     &&\ \ \ \ \ \ \  \ \   +  \Upsilon_{\beta}^{RA}(\epsilon_1-\omega_{\mathbf q_2}, \mathbf k')G^{A-}(\epsilon-\omega_{\mathbf q_2}, \mathbf k_2)( 1+n_B (\omega_{\mathbf q_2})- n_F(\epsilon_1-\omega_{\mathbf q_2}))] + ( M_X\leftrightarrow N_X, +\leftrightarrow -).
\end{eqnarray}

Since the phonon energy $\omega_{\mathbf q}$ is much smaller than the Fermi energy $\epsilon_F$, we ignore the phonon energy in all the electron GFs and $ \Upsilon_{\beta}^{RA}$, and only keep the phonon energy dependence in the distribution functions. Moreover, at low temperature, the phonon scattering processes are dominated by small phonon momentum scatterings. In the leading order of the phonon momentum, we have 
\begin{eqnarray}\label{eq:Simple_1}
 &&\ \ \ \delta(\epsilon_1 -\epsilon-\omega_{\mathbf q_1})[n_B(\omega_{\mathbf q_1})+n_F(\epsilon_1)] + \delta(\epsilon_1 -\epsilon+\omega_{\mathbf q_1})[1+n_B(\omega_{\mathbf q_1})-n_F(\epsilon_1)] \nonumber\\
    &&= [\delta(\epsilon_1 -\epsilon-\omega_{\mathbf q_1})n_B(\omega_{\mathbf q_1}) + \delta(\epsilon_1 -\epsilon+\omega_{\mathbf q_1})(1+n_B(\omega_{\mathbf q_1}))]\frac{1-n_F(\epsilon_1)}{1-n_F(\epsilon)} \nonumber\\
    &&\approx \frac{2\omega_{\mathbf q_1}}{k_BT}n_B(\omega_{\mathbf q_1})(1+n_B(\omega_{\mathbf q_1})) \delta(\epsilon_1-\epsilon),
\end{eqnarray}
and 
\begin{eqnarray}\label{eq:Simple_2}
&&\delta(\epsilon_2 -\epsilon_1-\omega_{\mathbf q_2})[n_B(\omega_{\mathbf q_2})+n_F(\epsilon_2)] + \delta(\epsilon_2 -\epsilon_1+\omega_{\mathbf q_2})[1+n_B(\omega_{\mathbf q_2})-n_F(\epsilon_2)] \nonumber\\
&\approx&  \frac{2\omega_{\mathbf q_2}}{k_BT}n_B(\omega_{\mathbf q_2})(1+n_B(\omega_{\mathbf q_2})) \delta(\epsilon_2-\epsilon_1).
\end{eqnarray}

Applying Eq.(\ref{eq:Simple_1}), (\ref{eq:Simple_2}) and  $ \Upsilon_{\beta}^{RA}(\epsilon, \mathbf k)\sim \delta(\epsilon-\epsilon_{\mathbf k})$, Eq.(\ref{eq:X_Response}) can be written as 
\begin{eqnarray}
\Pi_{\alpha\beta}^{X}(\omega\to 0)
    &=&  \frac{i\omega}{2\pi} \sum_{\mathbf k}\sum_{\mathbf k'}\sum_{\mathbf k_1}\sum_{\mathbf k_2} \delta(\mathbf k+\mathbf k'-\mathbf k_1-\mathbf k_2)
 W_{\mathbf k_1-\mathbf k} W_{\mathbf k_2 -\mathbf k} \int d\epsilon [-\partial_{\epsilon} n_F(\epsilon)]\Upsilon_{\alpha}^{AR}(\epsilon, \mathbf k) \Upsilon_{\beta}^{RA}(\epsilon, \mathbf k') \nonumber \\ 
    &&\times [ M_X(\mathbf k,\mathbf k_1,\mathbf k_2)G^{R+}(\epsilon,\mathbf k_1) G^{A-}(\epsilon,\mathbf k_2) + N_X(\mathbf k,\mathbf k_1,\mathbf k_2)G^{R-}(\epsilon,\mathbf k_1) G^{A+}(\epsilon,\mathbf k_2) ],
\end{eqnarray}
where $W_{\mathbf q}\equiv |g_{\mathbf q}|^2\frac{2\omega_{\mathbf q}}{k_BT}n_B(\omega_{\mathbf q})(1+n_B(\omega_{\mathbf q}))$.

Since 
\begin{equation}
 \Upsilon_{\alpha}^{AR}(\epsilon, \mathbf k) =G^{A+}(\epsilon,\mathbf k)\Gamma_\alpha^{++}(\mathbf k)G^{R+}(\epsilon,\mathbf k)\approx 2\pi\tau^{+}_k \delta(\epsilon-\epsilon_{k}^+) \Gamma_\alpha^{++}(\mathbf k),
\end{equation}
 and 
 $\Gamma_{\alpha}^{++}(\mathbf k) =\gamma j_\alpha^{++}(\mathbf k) =(\tau^{tr}_k/\tau_{k}^+) j_\alpha^{++}(\mathbf k)$, we have
\begin{equation}
  \Upsilon_{x}^{AR}(\epsilon,\epsilon;\mathbf k)=2\pi \tau^{tr}_k \delta(\epsilon-\epsilon_{k}^+) j_x^{++}(\mathbf k),  \ \ 
  \Upsilon_{y}^{RA}(\epsilon,\epsilon;\mathbf k')=2\pi \tau^{tr}_k \delta(\epsilon-\epsilon_{k'}^+) j_y^{++}(\mathbf k').
\end{equation}
The dc Hall conductivity from the $X$ diagram is then $\sigma^X_{xy}=\Pi_{\alpha\beta}^{X}(\omega\to 0)/i \omega$ is then
\begin{eqnarray}
 \sigma_{xy}^{X}
    &=&  2\pi \sum_{\mathbf k}\sum_{\mathbf k'}\sum_{\mathbf k_1}\sum_{\mathbf k_2} \delta(\mathbf k+\mathbf k'-\mathbf k_1-\mathbf k_2)
 W_{\mathbf k_1-\mathbf k} W_{\mathbf k_2 -\mathbf k} \int d\epsilon [-\partial_{\epsilon} n_F(\epsilon)] \delta(\epsilon-\epsilon_{k}^+)\delta(\epsilon-\epsilon_{k'}^+) (\tau^{tr}_k)^2 \nonumber \\ 
    && j_x^{++}(\mathbf k) j_y^{++}(\mathbf k')[ M_X(\mathbf k,\mathbf k_1,\mathbf k_2)G^{R+}(\epsilon,\mathbf k_1) G^{A-}(\epsilon,\mathbf k_2) + N_X(\mathbf k,\mathbf k_1,\mathbf k_2)G^{R-}(\epsilon,\mathbf k_1) G^{A+}(\epsilon,\mathbf k_2) ].
\end{eqnarray}

Denoting 
\begin{eqnarray}\label{eq:X_scattering_rate}
 \omega_{\mathbf k, \mathbf k'}^{X}
    &\equiv& 2\pi \delta(\epsilon_{k'}^+-\epsilon_k^+)\sum_{\mathbf k_1}\sum_{\mathbf k_2}\delta(\mathbf k+\mathbf k'-\mathbf k_1-\mathbf k_2)W_{\mathbf k_1-\mathbf k} W_{\mathbf k_2-\mathbf k} \nonumber\\
 && [ M_X(\mathbf k,\mathbf k_1,\mathbf k_2)G^{R+}(\epsilon,\mathbf k_1) G^{A-}(\epsilon,\mathbf k_2) + N_X(\mathbf k,\mathbf k_1,\mathbf k_2)G^{R-}(\epsilon,\mathbf k_1) G^{A+}(\epsilon,\mathbf k_2) ],
\end{eqnarray}
which corresponds to the scattering rate from state $\mathbf k$ to $\mathbf k'$ through the two phonon scattering process in the $X$ diagram, the AH conductivity $ \sigma_{xy}^{X}$ can be written as 
\begin{eqnarray}
 \sigma_{xy}^{X}\label{eq:X_tau_per}
    = \sum_{\mathbf k}\sum_{\mathbf k'}\omega_{\mathbf k, \mathbf k'}^{X}
 \int d\epsilon [-\partial_{\epsilon} n_F(\epsilon)] \delta(\epsilon-\epsilon_{k}^+)
    (\tau^{tr}_k)^2 j_x^{++}(\mathbf k) j_y^{++}(\mathbf k').
\end{eqnarray}

The AH conductivity corresponds to the anti-symmetric part of $ \sigma_{xy}^{X}$:
\begin{equation}\label{eq:X_AH_conductivity}
\sigma_{xy}^{X,a}
    =\frac{1}{2} \sum_{\mathbf k}\sum_{\mathbf  k'}\omega_{\mathbf  k, \mathbf  k'}^{X}
 \int d\epsilon [-\partial_{\epsilon} n_F(\epsilon)] \delta(\epsilon-\epsilon_{k}^+)
    (\tau^{tr}_k)^2 [j_x^{++}(\mathbf  k) j_y^{++}(\mathbf  k') -j_y^{++}(\mathbf  k) j_x^{++}(\mathbf  k')].
\end{equation}

Since $j_x^{++}(\mathbf k)= ev \sin\alpha_0 \cos\theta_0,\  j_y^{++}(\mathbf k')=  ev \sin\alpha'_0 \sin(\theta_0+\theta)$, where $\sin \alpha'_0\approx \sin\alpha_0= vk/\epsilon_k$, 
\begin{equation}
\sigma_{xy}^{X,a}
    = \frac{1}{2}e^2v^2 \sum_{\mathbf k}\sum_{\mathbf k'}\omega_{\mathbf k, \mathbf k'}^{X}
 \int d\epsilon [-\partial_{\epsilon} n_F(\epsilon)] \delta(\epsilon-\epsilon_{k}^+)(\tau^{tr}_k)^2  \sin^2{\alpha_0}\sin\theta,
\end{equation}
where $\theta=\phi_{\mathbf k'}- \phi_{\mathbf k}$ is the angle between $\mathbf k$ and $\mathbf k'$. Denoting
\begin{equation}
  \frac{1}{\tau^{\perp}_{k,X}}\equiv - \sum_{\mathbf k'} \omega_{\mathbf k, \mathbf k'} ^X\sin\theta,
\end{equation}
the AH conductivity for the $X$ diagram can be written as 
\begin{eqnarray}
\sigma_{xy}^{X,a}
    &=& -\frac{1}{2}e^2v^2 \sum_{\mathbf k}\frac{(\tau^{tr}_k)^2 }{\tau^{\perp}_{k,X}}\int d\epsilon [-\partial_{\epsilon} n_F(\epsilon)] \delta(\epsilon-\epsilon_{k}^+)  \sin^2{\alpha_0} \nonumber\\
    &=& -e^2v^2 \frac{k_F^2}{4\pi \epsilon_F} \frac{(\tau^{tr}_{k_F})^2 }{\tau^{\perp}_{k_F, X}}.
\end{eqnarray}
This is consistent with the general result of the skew scattering contribution obtained from the semi-classical Boltzmann equation approach in Ref.\cite{Sinitsyn2007}.

Similarly, we can get the AH conductivity from the $\Psi$ diagrams as 
\begin{eqnarray}
\sigma_{xy}^{\Psi,a}
    &=& -e^2v^2 \frac{k_F^2}{4\pi \epsilon_F} \frac{(\tau^{tr}_{k_F})^2 }{\tau^{\perp}_{k_F, \Psi}},
\end{eqnarray}
where 
\begin{equation}\label{eq:Psi_tau_per}
  \frac{1}{\tau^{\perp}_{k, \Psi}}\equiv -\sum_{\mathbf k'} \omega_{\mathbf k, \mathbf k'} ^\Psi\sin\theta,
\end{equation}
and  $\omega_{\mathbf k, \mathbf k'} ^\Psi$ is the scattering rate from $\mathbf k$ to $\mathbf k'$ for the $\Psi$ diagrams.

The leading order $\Psi$ diagrams in the chiral basis are shown in Fig.\ref{fig:crossed_diagrams}(c)-(f). Similar to the $X$ diagrams, we obtain the scattering rates from $\mathbf k$ to $\mathbf k'$ for these diagrams as 
\begin{eqnarray}
 \omega^{\Psi,(c)}_{\mathbf k, \mathbf k'}
    &= & 2\pi \delta(\epsilon_{k'}^+-\epsilon_k^+)\sum_{\mathbf k_1}\sum_{\mathbf k_2}\delta(\mathbf k'+\mathbf k_1-\mathbf k-\mathbf k_2)  
    W_{\mathbf k-\mathbf k_1} W_{\mathbf k'-\mathbf k} G^{R+}(\mathbf k_1) G^{A-}(\mathbf k_2) \inner{u_{\mathbf k}^+}{u_{\mathbf k_1}^+} \inner{u_{\mathbf k_1}^+}{u_{\mathbf k_2}^-}\inner{u_{\mathbf k_2}^-}{u_{\mathbf k'}^+}\inner{u_{\mathbf k'}^+}{u_{\mathbf k}^+}, \nonumber\\
   \omega^{\Psi,(d)}_{\mathbf k, \mathbf k'}
    &=& 2\pi \delta(\epsilon_{k'}^+-\epsilon_k^+)\sum_{\mathbf k_1}\sum_{\mathbf k_2}\delta(\mathbf k'+\mathbf k_1-\mathbf k-\mathbf k_2) 
     W_{\mathbf k-\mathbf k_1} W_{\mathbf k'-\mathbf k} G^{A-}(\mathbf k_1) G^{R+}(\mathbf k_2)\inner{u_{\mathbf k}^+}{u_{\mathbf k_1}^-} \inner{u_{\mathbf k_1}^-}{u_{\mathbf k_2}^+}\inner{u_{\mathbf k_2}^+}{u_{\mathbf k'}^+}\inner{u_{\mathbf k'}^+}{u_{\mathbf k}^+}, \nonumber
\end{eqnarray}
and $\omega^{\Psi,(e)}_{\mathbf k, \mathbf k'}=[\omega^{\Psi,(c)}_{\mathbf k, \mathbf k'}]^*,
\omega^{\Psi,(f)}_{\mathbf k, \mathbf k'}=[\omega^{\Psi,(d)}_{\mathbf k, \mathbf k'}]^*$.

The total scattering rate of all the $\Psi$ diagrams is then
\begin{eqnarray}\label{eq:Psi_scattering_rate}
 \omega^{\Psi}_{\mathbf k, \mathbf k'} &=& 2\pi \delta(\epsilon_{k'}^+-\epsilon_k^+)\sum_{\mathbf k_1}\sum_{\mathbf k_2}  W_{\mathbf k'-\mathbf k} G^{R+}(\mathbf k_1)G^{A-}(\mathbf k_2) \nonumber\\
   & & [\delta(\mathbf k'+\mathbf k_1-\mathbf k-\mathbf k_2)W_{\mathbf k-\mathbf k_1}  M_{\Psi}(\mathbf k, \mathbf k_1, \mathbf k_2)
    + \delta(\mathbf k'+\mathbf k_2-\mathbf k-\mathbf k_1) W_{\mathbf k-\mathbf k_2} N_{\Psi}(\mathbf k, \mathbf k_1, \mathbf k_2)  ] +c.c., 
\end{eqnarray}
where 
\begin{equation}
 M_{\Psi}(\mathbf k, \mathbf  k_1, \mathbf  k_2)\equiv \inner{u_{\mathbf  k}^+}{u_{\mathbf  k_1}^+} \inner{u_{\mathbf  k_1}^+}{u_{\mathbf  k_2}^-}\inner{u_{\mathbf  k_2}^-}{u_{\mathbf  k'}^+}\inner{u_{\mathbf  k'}^+}{u_{\mathbf  k}^+},
 \  N_{\Psi}(\mathbf  k, \mathbf  k_1, \mathbf  k_2)\equiv \inner{u_{\mathbf  k}^+}{u_{\mathbf  k_2}^-} \inner{u_{\mathbf  k_2}^-}{u_{\mathbf  k_1}^+}\inner{u_{\mathbf  k_1}^+}{u_{\mathbf  k'}^+}\inner{u_{\mathbf  k'}^+}{u_{\mathbf  k}^+}.
\end{equation}

Since $\tau^{tr}_{k_F}$ is given in Appendix B, the key ingredient to obtain the AH conductivities $\sigma_{xy}^{X,a}$ and $\sigma_{xy}^{\Psi,a}$ is to calculate  $1/\tau^{\perp}_{k, X}$ and  $1/\tau^{\perp}_{k, \Psi}$ or  $\omega^{X}_{\mathbf k, \mathbf k'} $ and  $\omega^{\Psi}_{\mathbf k, \mathbf k'} $ in Eq.(\ref{eq:X_scattering_rate}) and Eq.(\ref{eq:Psi_scattering_rate}). From Eq.(\ref{eq:X_AH_conductivity}), one can see that only the anti-symmetric part of $\omega^{X}_{\mathbf k, \mathbf k'}$ contributes to the AH conductivity, so does $\omega^{\Psi}_{\mathbf k, \mathbf k'}$. From Eq.(\ref{eq:X_scattering_rate}) and Eq.(\ref{eq:Psi_scattering_rate}), we get the anti-symmetric part of the scattering rates of the $X$ and $\Psi$ diagrams as
\begin{eqnarray}
 \omega^{X,a}_{\mathbf k, \mathbf k'} &=&  4\pi^2 \delta(\epsilon_{k'}^+-\epsilon_k^+)\sum_{\mathbf k_1}\sum_{\mathbf k_2}\delta(\mathbf k+\mathbf k'-\mathbf k_1-\mathbf k_2)W_{\mathbf k-\mathbf k_1} W_{\mathbf k'-\mathbf k_1} \frac{\delta(\epsilon-\epsilon_{\mathbf k_1}^+)}{\epsilon-\epsilon_{\mathbf k_2}^{-}}\text{Im}(\inner{u_{\mathbf k}^+}{u_{\mathbf k_1}^+}\inner{u_{\mathbf k_1}^+}{u_{\mathbf k'}^+}\inner{u_{\mathbf k'}^+}{u_{\mathbf k_2}^-} \inner{u_{\mathbf k_2}^-}{u_{\mathbf k}^+}) , \nonumber\\
\omega^{\Psi,a}_{\mathbf k, \mathbf k'}&=& 4\pi^2  \delta(\epsilon_{k'}^+-\epsilon_k^+)\sum_{\mathbf k_1}\sum_{\mathbf k_2}\delta(\mathbf k'+\mathbf k_1-\mathbf k-\mathbf k_2) 
     W_{\mathbf k-\mathbf k_1} W_{\mathbf k'-\mathbf k}
     \left[ \frac{\delta(\epsilon-\epsilon_{\mathbf k_1}^+)}{\epsilon-\epsilon_{\mathbf k_2}^{-}}\text{Im}(\inner{u_{\mathbf k}^+}{u_{\mathbf k_1}^+} \inner{u_{\mathbf k_1}^+}{u_{\mathbf k_2}^-}\inner{u_{\mathbf k_2}^-}{u_{\mathbf k'}^+}\inner{u_{\mathbf k'}^+}{u_{\mathbf k}^+}) \right. \nonumber\\
  &&\left. \ \ \ \ \ \ \ \ \ \ \ \ \ \ \ \ \ \ \  \ \ \ \ \ \ \ \ \ \ \ \ \ \ \ +   \frac{\delta(\epsilon-\epsilon_{\mathbf k_2}^+)}{\epsilon-\epsilon_{\mathbf k_1}^{-}} \text{Im}(\inner{u_{\mathbf k}^+}{u_{\mathbf k_1}^-} \inner{u_{\mathbf k_1}^-}{u_{\mathbf k_2}^+}\inner{u_{\mathbf k_2}^+}{u_{\mathbf k'}^+}\inner{u_{\mathbf k'}^+}{u_{\mathbf k}^+})\right].
\end{eqnarray}

It is hard to compute  $\omega^{X,a}_{\mathbf k, \mathbf k'}$ and $\omega^{\Psi,a}_{\mathbf k, \mathbf k'} $ in the whole temperature regime. But at low temperature, the phonon momenta participating in the scatterings are small and we can expand $\omega^{X,a}_{\mathbf k, \mathbf k'}$ and $\omega^{\Psi,a}_{\mathbf k, \mathbf k'} $ in terms of the phonon momentum. At low temperature
\begin{equation}
 \inner{u_{\mathbf k}^s}{u_{\mathbf k'}^{s'}}\approx \inner{u_{\mathbf k}^s}{u_{\mathbf k}^{s'}} + (\mathbf k' -\mathbf k)\cdot \inner{u_{\mathbf k}^s}{ \nabla _{\mathbf k} u_{\mathbf k}^{s'}}=\delta_{s,s'}+   \mathbf {\mathcal{A}}^{ss'}(\mathbf k)\cdot (\mathbf k' -\mathbf k), \ s,s'=\pm,
\end{equation}
where ${\mathcal{A}}^{ss'}(\mathbf k)\equiv\inner{u_{\mathbf k}^s}{ \nabla _{\mathbf k} u_{\mathbf k}^{s'}}$ is the Berry connection. In the leading order of the expansion, we have 
\begin{eqnarray}
 \text{Im}(\inner{u_{\mathbf k}^+}{u_{\mathbf  k_1}^+}\inner{u_{\mathbf  k_1}^+}{u_{\mathbf  k'}^+}\inner{u_{\mathbf  k'}^+}{u_{\mathbf k_2}^-} \inner{u_{\mathbf  k_2}^-}{u_{\mathbf  k}^+}) &\approx&
  (\mathbf  k_2 -\mathbf  k')_{\alpha}(\mathbf  k-\mathbf  k_2)_{\beta}\text{Im}( \mathcal{A}_{\alpha}^{+-}(\mathbf  k)  \mathcal{A}_{\beta}^{-+}(\mathbf  k) ), \\
 \text{Im}(\inner{u_{\mathbf k}^+}{u_{\mathbf k_1}^+} \inner{u_{\mathbf k_1}^+}{u_{\mathbf k_2}^-}\inner{u_{\mathbf k_2}^-}{u_{\mathbf k'}^+}\inner{u_{\mathbf k'}^+}{u_{\mathbf k}^+})&\approx&
 (\mathbf k_2-\mathbf k_1)_{\alpha}(\mathbf k'-\mathbf k_2)_{\beta}  \text{Im}(\mathcal{A}^{+-}_{\alpha}(\mathbf k)   \mathcal{A}^{-+}_{\beta}(\mathbf k) ), \\
 \text{Im}(\inner{u_{\mathbf k}^+}{u_{\mathbf k_1}^-} \inner{u_{\mathbf k_1}^-}{u_{\mathbf k_2}^+}\inner{u_{\mathbf k_2}^+}{u_{\mathbf k'}^+}\inner{u_{\mathbf k'}^+}{u_{\mathbf k}^+})&\approx&
 (\mathbf k_2-\mathbf k_1)_{\alpha}(\mathbf k_1-\mathbf k)_{\beta}  \text{Im}(\mathcal{A}^{-+}_{\alpha}(\mathbf k)   \mathcal{A}^{+-}_{\beta}(\mathbf k)),
\end{eqnarray}
where sum over repeated indices is implied.

In the leading order of phonon momentum expansion, $\omega^{X,a}_{\mathbf k, \mathbf k'}$ and $\omega^{\Psi ,a}_{\mathbf k, \mathbf k'}$ then become  
\begin{eqnarray}
\omega^{X,a}_{\mathbf k, \mathbf k'}&\approx &
4\pi^2 \frac{\delta(\epsilon_{k'}^+-\epsilon_k^+)}{\epsilon+\epsilon^+_{k}}\text{Im}[\mathcal{A}_{\alpha}^{+-}(\mathbf k)  \mathcal{A}_{\beta}^{-+}(\mathbf k) ]\sum_{\mathbf k_1} 
     W_{\mathbf k-\mathbf k_1} W_{\mathbf k'-\mathbf k_1} \delta(\epsilon-\epsilon_{\mathbf k_1}^+)(\mathbf k -\mathbf k_1)_{\alpha}(\mathbf k_1-\mathbf k')_{\beta}, \label{eq:X_scattering_expansion}\\
 \omega^{\Psi, a}_{\mathbf k, \mathbf k'}&\approx & 4\pi^2  \delta(\epsilon_{k'}^+-\epsilon_k^+)\text{Im}[\mathcal{A}^{-+}_{\alpha}(\mathbf k)   \mathcal{A}^{+-}_{\beta}(\mathbf k)]
 \left[\sum_{\mathbf k_1} 
     W_{\mathbf k-\mathbf k_1} W_{\mathbf k'-\mathbf k} (\mathbf k'-\mathbf k)_{\alpha}(\mathbf k_1-\mathbf k)_{\beta} 
      \frac{\delta(\epsilon-\epsilon_{\mathbf k_1}^+)}{\epsilon-\epsilon_{\mathbf k'+\mathbf k_1-\mathbf k}^{-}} \right. \nonumber\\
     &&\left. +  \sum_{\mathbf k_2}
     W_{\mathbf k'-\mathbf k_2} W_{\mathbf k'-\mathbf k} (\mathbf k'-\mathbf k)_{\alpha}(\mathbf k_2-\mathbf k')_{\beta} \frac{\delta(\epsilon-\epsilon_{\mathbf k_2}^+)}{\epsilon-\epsilon_{\mathbf k+\mathbf k_2-\mathbf k'}^{-}} \right] \nonumber\\
     &\approx& 8\pi^2   \frac{\delta(\epsilon_{k'}^+-\epsilon_k^+)}{\epsilon+\epsilon_{k}^+}\text{Im}[\mathcal{A}^{+-}_{\alpha}(\mathbf k)   \mathcal{A}^{-+}_{\beta}(\mathbf k)] \sum_{\mathbf k''}
    W_{\mathbf k-\mathbf k''} W_{\mathbf k'-\mathbf k} (\mathbf k-\mathbf k')_{\alpha}(\mathbf k''-\mathbf k)_{\beta} \delta(\epsilon-\epsilon_{\mathbf k''}^+). \label{eq:Psi_scattering_expansion}
\end{eqnarray}
We have applied the quasi-elastic approximation for the e-phonon scatterings near the Fermi surface and $\epsilon_{\mathbf k}^-=-\epsilon_{\mathbf k}^+$ in the above equations.

The leading order expansion of $\omega^{X,a}_{\mathbf k, \mathbf k'}$ in Eq.(\ref{eq:X_scattering_expansion}) is exactly opposite to the leading order  of the scattering rate of the non-crossing skew scattering diagrams Fig.\ref{fig:skew_scattering}(a)+(b), which is shown in Eq.(39) of the supplementary of Ref.\cite{Xiao2019-1}.  Similarly, the leading order of $\omega^{\Psi,a}_{\mathbf k, \mathbf k'}$ in Eq.(\ref{eq:Psi_scattering_expansion}) is exactly opposite to 
the leading order expansion of the total scattering rate of the diagrams Fig.\ref{fig:skew_scattering}(c)-(f), which is shown in Eq.(37) and (42) of the supplementary of Ref.\cite{Xiao2019-1}. From Eq.(\ref{eq:X_tau_per})-(\ref{eq:Psi_tau_per}), we can see that at low temperature,  the AH conductivity of the crossed $X$ and $\Psi$ diagrams is exactly opposite to that of the non-crossing skew scattering contribution in the leading order of the temperature. We have obtained the latter, i.e., the intrinsic skew scattering contribution in the last subsection of this work and checked that it is consistent with that obtained in the semi-classical approach in Ref.\cite{Xiao2019-1}.

\section{Appendix D: Screening effect}
We discuss the screening effect due to e-e interaction in this section. To take into account this effects, we add the Thomas-Fermi (TF) screening factor to the deformation potential, i.e., we replace $g_D$ by  $g_D\frac{q}{q+q_{TF}}$ where $q_{TF}=\alpha \epsilon_F/v$ is the TF wave-vector for 2D massive Dirac metals and $\alpha=e^2/\hbar v$ is the fine structure constant. With this replacement, the only change we need to make in the calculation of the AH conductivity is to replace the parameters $a, b, c$ by
 \begin{eqnarray}
 a^{sc}(t)&=&C\int_0^{1/t} x dx \left(\frac{x}{x+\frac{\alpha}{2t}\frac{1}{\sqrt{1-\Delta^2/\epsilon^2_F}}}\right)^2(1- t^2 x^2)^{-\frac{1}{2}}\left(\frac{1}{e^x-1}+\frac{1}{e^x+1}\right), \\
 b^{sc}(t)&=&C\int_0^{1/t} x dx \left(\frac{x}{x+\frac{\alpha}{2t}\frac{1}{\sqrt{1-\Delta^2/\epsilon^2_F}}}\right)^2(1-2t^2 x^2)(1- t^2 x^2)^{-\frac{1}{2}}\left(\frac{1}{e^x-1}+\frac{1}{e^x+1}\right), \\
   c^{sc}(t)&=&C\int_0^{1/t} x dx \left(\frac{x}{x+\frac{\alpha}{2t}\frac{1}{\sqrt{1-\Delta^2/\epsilon^2_F}}}\right)^2(1-2t^2 x^2)^2(1- t^2 x^2)^{-\frac{1}{2}}\left(\frac{1}{e^x-1}+\frac{1}{e^x+1}\right).
 \end{eqnarray}
 From the above equations, we can see that $a^{sc}(t), b^{sc}(t), c^{sc}(t)$ not only depend on $\Delta/\epsilon_F$ and $t\equiv T/T_{BG}$, but also depend on $\alpha$ in general.

The modification of $a, b, c$ by the screening results in a modification of the AH conductivities. 
In Fig.\ref{fig:AH_conductivity} of the main text, we show the difference of the AH conductivities with and without screening as a function of the rescaled temperature $t\equiv T/T_{BG}$. 
In the low temperature limit $T\ll T_{BG}$, we can expand $a^{sc}(t), b^{sc}(t), c^{sc}(t)$ and get 
\begin{eqnarray}
a^{sc}(t)& \approx &\kappa \left(1+\pi^2 \frac{T^2}{T^2_{BG}}+\frac{51}{16} \pi^4 \frac{T^4}{T^2_{BG}}\right),  \\
b^{sc}(t)& \approx &\kappa \left(1-3\pi^2 \frac{T^2}{T^2_{BG}}-\frac{85}{16} \pi^4 \frac{T^4}{T^2_{BG}}\right),  \\
c^{sc}(t)& \approx &\kappa \left(1-7\pi^2 \frac{T^2}{T^2_{BG}}+\frac{323}{16} \pi^4 \frac{T^4}{T^2_{BG}}\right),
\end{eqnarray}
where $\kappa\equiv\frac{\pi^2}{2\alpha^2}\frac{ T^2}{T^2_{BG}}(1-\Delta^2/\epsilon^2_F). $

The AH conductivities at $T\ll T_{BG}$ with screening are then
\begin{eqnarray}
\tilde{\sigma}^{side}_{xy}(\frac{\Delta}{\epsilon_F}, t)&\approx & -\frac{e^2}{4\pi }\frac{\Delta}{\epsilon_F} (1-\frac{\Delta^2}{\epsilon_F^2}) \left[1+\frac{17}{4}\pi^2(1-\frac{\Delta^2}{\epsilon^2_F})\frac{T^2}{T^2_{BG}}\right],  \label{eq:side_sc}\\
\tilde{\sigma}^{sk-nc}_{xy}(\frac{\Delta}{\epsilon_F}, t) &\approx &-\frac{17}{16}\pi e^2\frac{\Delta}{\epsilon_F} (1-\frac{\Delta^2}{\epsilon_F^2})^2 \frac{T^2}{T^2_{BG}}. \label{eq:sk_sc}
\end{eqnarray}
Comparing Eq.(\ref{eq:side_sc}) and (\ref{eq:sk_sc}) with the AH conductivities without screening in Table I of the main text, we can see that the AH conductivities in the limit $T\to0$ are not changed by the screening. But at finite temperature, the screening modifies the AH conductivities as shown in the plots of Fig.2 in the main text.

 At the high temperature limit $T\gg T_{BG}$, the AH conductivities depend on the TF wave vector and there is no simple analytical result for the screened case.

\end{widetext}

\end{document}